\documentclass{aa}
\usepackage{psfig}

\def\solar {\ifmmode_{\mathord\odot} \else $_{\mathord\odot}$\fi} 
\def\Msol {\ifmmode {\,{\it M}\solar} \else $\,M$\solar\fi}        

\begin{document}
\hyphenation{ELO-DIE Leven-berg ra-dial mas-ses Del-fosse}
   \thesaurus{06     
              (08.02.3;  
              08.02.4;  
              08.02.6;  
              08.12.2  
              03.20.7;  
              08.12.1)  
    }

   \title{Accurate masses of very low mass stars:\\
  III 16 new or improved masses.
 \thanks{Based on observations made at the Observatoire de Haute 
   Provence (CNRS), and at the CFH Telescope, operated by the NRCC, the CNRS
   and the University of Hawaii. Tables 4-15 are available only in 
   electronic form at the CDS via anonymous ftp
   to cdsarc.u-strasbg.fr.(130.79.128.5) or via http://cdsweb.u-strasbg.fr/Abstract.html }
}
\authorrunning{S\'egransan et al.}
\titlerunning{16 accurate M dwarf masses} 
   \author{D.~S\'egransan \inst{1}, X.~Delfosse \inst{2}, 
T.~Forveille \inst{1,3}, J.-L Beuzit \inst{1,3}, S.~Udry \inst{4}, 
C.~Perrier \inst{1} and M.~Mayor \inst{4}
          }

   \offprints{Damien S\'egransan, e-mail: segransa@obs.ujf-grenoble.fr}

\institute{   Observatoire de Grenoble,
              414 rue de la Piscine,
              Domaine Universitaire de S$^{\mathrm t}$ Martin d'H\`eres,
              F-38041 Grenoble,
              France
\and
              Instituto de Astrofisica de Canarias
              E-38200 La Laguna, Tenerife, 
              Canary Islands, Spain
\and
              Canada-France-Hawaii Telescope Corporation, 
              P.O. Box 1597,
              Kamuela, HI 96743, 
              U.S.A.
\and
              Observatoire de Gen\`eve,
              51 Ch des Maillettes,
              1290 Sauverny,
              Switzerland   
}

   \date{Received 21 July 2000, Accepted 28 September 2000}

   \maketitle

   \begin{abstract}

We have obtained adaptive optics images and accurate radial velocities 
for 7 very low mass objects, In the course of a long term effort to 
determine accurate masses for very low mass stars (M$<$0.6~\Msol).
We use the new data, together with measurements from 
the litterature for some stars, to determine new or improved orbits 
for these 7 systems. They provide masses for 16 very low mass
stars with accuracies that range between 0.2\% and 5\%, and in some cases
a very accurate distance as well. This information is used in a companion 
paper to discuss the Mass-Luminosity relation for the V, J, H and K 
photometric bands.

   \end{abstract}

%

\section{Introduction}
Accurate masses for binary stars provide a crucial test of our understanding 
of stellar physics (e.g. Andersen \cite{andersen91}). Mass, the basic 
input of evolutionary models, is directly
measured, and the models must reproduce the effective temperatures and
luminosities (or radii) of both components, for a single age and a
single chemical composition. Given the strong mass dependency of all
stellar parameters however, this discriminating diagnostic 
only shows its power for relative mass errors $\leq$1-3\%. 

Very accurate mass measurements have long been the exclusive
province of double-lined detached eclipsing binaries
(Andersen \cite{andersen91}, \cite{andersen98}), for which as a bonus the 
stellar radii are simultaneously determined. Such systems are however 
unfortunately rare,
and only 44 pairs had yielded masses accurate enough to be included in 
Andersen \cite{andersen91}'s critical compilation, mostly for 
intermediate mass stars.
Relatively few eclipsing binaries have had their masses measured since then. 
In the mass range of interest here, the litterature still contains
no more than three well detached eclipsing binaries with substantially
subsolar component masses: YY~Gem (M0Ve, 0.6+0.6\Msol; Bopp \cite{bopp74}; 
Leung \& Schneider \cite{leung78}), the recently identified GJ~2069A 
(M3.5Ve, 0.4+0.4\Msol; Delfosse et al. \cite{delfosse99a}), 
and CM Dra (M4Ve, 0.2+0.2\Msol; Lacy \cite{lacy77}; 
Metcalfe et al. \cite{metcalfe96}). 

Angularly resolved spectroscopic binaries provide stellar masses in
parts of the HR diagram where eclipsing systems are rare or missing,
and in particular for very low mass stars.
Until recently however, these measurements did not match the $\sim$1\% 
accuracy which can be obtained in detached eclipsing systems. As a
consequence, the best representation to date of the empirical M-L relation 
for M dwarfs had to mostly rely on masses determined with 5-20\% accuracy 
(Henry \& McCarthy \cite{henry93}; Henry et al. \cite{henry99}). 
The last two years have seen a 
dramatic evolution in this respect, with two groups breaking through
the former $\sim$5\% accuracy barrier. The first group to do so used the 
1~mas per measurement astrometric  accuracy of the Fine Guidance Sensors 
(Benedict et al. \cite{benedict99}) on {\it HST} to determine a few 
masses of angularly resolved binaries with 2 to 10\% accuracy (Franz et 
al. \cite{franz98}; Torres et al. \cite{torres99}; Henry et al. 
\cite{henry99}; Benedict et al. \cite{benedict00}). Slightly more 
recently, we have demonstrated that
the combination of very accurate radial velocities with angular
separations from adaptive optics imaging can yield masses for
VLMS with even better accuracy, of only 1-3\% 
(Forveille et al. \cite{forveille99}; Delfosse et al. \cite{delfosse99b}). 
Here we present 16 new or improved masses
determined with the same method, with accuracies that now range
between 0.2 and 5\%. In a companion paper (Delfosse et al. 
\cite{delfosse00a}) we 
rediscuss the VLMS mass-luminosity relation in the light of these
new data.
We first discuss the observing program and its sample in Sect.~2 and then 
present the observations and data processing in Sect.~3. Section~4 describes
the orbit adjustment and the mass determination.

\section{Observing program }

Since 1995 we observe a distance-limited-sample of 
solar neighbourhood M dwarfs with adaptive optics imaging
and high accuracy radial-velocity monitoring (Delfosse et al. 
\cite{delfosse99c},
for a full presentation of the project). This long term effort
has two main motivations: 
\begin{itemize} 
\item{} determine the multiplicity statistics of disk M dwarfs with 
negligible incompleteness corrections, 
\item{} measure very accurate stellar masses.
\end{itemize}
To specifically address
the second goal, we have complemented the main distance limited sample
with a few well known M dwarfs binaries, beyond its distance or declination
limits. These additional binaries have periods longer than $\sim$1 year
(shorter period systems beyond a few parsecs are unresolved 
by 4~meter-class telescopes in the near-IR) and shorter than 
$\sim$10~years (to limit the timescale of the program).

\section{Observations and data processing}

\subsection{Radial-velocity observations}
We measure radial-velocities for stars in our sample with the \\ELODIE
spectrograph (Baranne et al. \cite{baranne96}) on the 1.93~m telescope of the
Observatoire de Haute Provence (France). 
This fixed configuration dual-fiber-fed echelle
spectrograph covers in a single exposure the 390-680~nm spectral
range, at an average resolving power of 42000. An elaborate on-line 
processing is integrated with the
spectrograph control software, and automatically produces optimally
extracted and wavelength calibrated spectra, with algorithms described
in Baranne et al. (\cite{baranne96}). All stars in this programme 
are observed with
a Thorium lamp illuminating the monitoring fiber, as needed for the best
($\sim 10~{\rm m\,s^{-1}}$) radial-velocity precision. 
The present paper uses data obtained between September 1995 and April 2000. 

A few measurements were also obtained with the CORALIE spectrograph on the
recently commissioned 1.2-m Euler telescope at La Silla Observatory
(Chile). CORALIE is an improved copy of ELODIE and has very similar
characteristics, with the exception of a substantially improved intrinsic
stability and a somewhat higher spectral resolution ($R=50000$). 

These spectra are analysed for velocity by numerical cross-correlation 
with a one-bit (i.e. 0/1) template. This processing is standard for ELODIE 
spectra (Queloz \cite{queloz95a}, \cite{queloz95b}). The correlation mask 
used here was derived by Delfosse et al. (\cite{delfosse99c}) from a high 
S/N spectrum of Gl~699 (Barnard's stars, M4V). As discussed 
in Sect.~4, we determine the orbital parameters of double-lined systems 
through a direct least square adjustment to the correlation profiles. We
recommend that reanalyses of those data similarly use those profiles
(available upon request to the authors). For easier reference we
nonetheless provide in Table~\ref{ind_meas1} to \ref{ind_meas7} 
(only available in the electronic version of this paper) 
radial velocities 
for all stars, obtained from adjustment of Gaussian functions to the 
correlation profiles. The measurement accuracies 
for the sources discussed here range between 10 and 100 ${\rm m\,s^{-1}}$ 
(depending on apparent magnitude and spectral 
type), except for the two fastest rotators, GJ~2069~A and YY~Gem.

\subsection{Adaptive optics imaging}
Adaptive optics observations are obtained at the 3.6-meter 
Canada-France-Hawaii
Telescope (CFHT) using PUE'O, the CFHT Adaptive Optics Bonnette (Arsenault
et al. \cite{arsenault94}, 
Rigaut et al. \cite{rigaut98}) and two different infrared 
cameras (Nadeau et al. \cite{nadeau94}, 
Doyon et al. \cite{doyon98}). Delfosse et al. (\cite{delfosse99c}) 
provide a detailed
description of the observing procedure, which we 
only summarize here.


The program stars are observed in a 4 or 5 positions mosaic pattern, that
allows to both determine the sky background from the on-source frames and 
fully compensate the cosmetic defects of the detector. The science targets 
are used to sense and correct the incoming wavefront. All of them are 
bright enough (R~$<$~14) to ensure diffraction-limited images in the H
and K bands under standard Mauna Kea atmospheric conditions (i.e. 
for seeing up to 1\arcsec). The corrected point spread function obtained
from the AO system is synthesized from simultaneous records of the
wavefront sensor measurements and deformable mirror commands, as described
by V\'eran et al. (\cite{veran97}). For pre-1997 observations this ancillary
information was not available from the acquisition system, and the point
spread function is then instead estimated 
from observations of a reference single star of similar R-band magnitude.
Astrometric calibration fields such as the central region of
the Trapezium Cluster in the Orion Nebula (Mc Caughrean \& Stauffer 
\cite{mccaughrean94}),
were observed to accurately determine the actual detector plate scale
and orientation on the sky.

In good seeing conditions the binaries
are observed through J (1.2 $\mu$m), H (1.65 $\mu$m) and 
K(2.23 $\mu$m) filters, or through corresponding narrow-band 
filters (usually [Fe$^+$] (1.65 $\mu$m) and Br$\gamma$ (2.166 $\mu$m))
for sources which would otherwise saturate the detectors in the minimum 
available integration time. 
For worse seeing we restrict observations to the K band, 
to maintain an acceptable corrected image quality.

We use a deconvolution algorithm (V\'eran et al. \cite{veran99}) based 
on  the Levenberg-Marquardt minimisation method and coded within IDL
to determine the separation, position angle and magnitude difference 
between the two stars. With approximate initial values of the positions of
the two components along with the PSF reference image, the fitting
procedures outputs the flux and pixel coordinates of both stars.
The astrometric calibrations then yields the desired angular 
separations. Table~\ref{ind_meas8} to \ref{ind_meas12} 
(only available electronically) list the individual measurements.

Additional angular separations could be obtained from 
the litterature for some binaries. They are also  listed in 
Table~\ref{ind_meas8} to \ref{ind_meas12}, and discussed in 
Sect.~\ref{individual} for each relevant system


\subsection{Parallaxes}
As discussed in Sect.~4, the orbital adjustment can make use of the 
trigonometric parallax of a multiple system, which is handled as an 
additional observational constraint on the ratio of its physical and 
angular dimensions. We have obtained this information (Table~\ref{photo})
from the Yale  General Catalog of trigonometric Parallaxes (Van Altena et al.,
\cite{vanaltena95}) and the HIPPARCOS 
catalog (ESA \cite{esa97}), with some individual entries from 
Probst (\cite{probst77})
and Soderhjelm (\cite{soderhjelm99}).

\begin{table}
\tabcolsep 1.15mm
\begin{tabular}{|l|llll|} \hline
Name     & Hipparcos       & Yale          & Soderhjelm      & Probst        \\ \hline
Gl~234   &                 &               & 244.2$\pm$2.4   & 243.2$\pm$2.0 \\
YY Gem   &                 & 74.7$\pm$2.5  &                 &               \\
GJ 2069A & 78.05$\pm$5.69  &               &                 &               \\
Gl~644   & 153.96$\pm$4.04 & 154.8$\pm$0.6 & 155.63$\pm$1.81 &               \\
Gl~747   &                 & 122.3$\pm$2.5 &                 &               \\
Gl~831   & 124.82$\pm$2.88 & 125.8$\pm$2.3 &                 &               \\
Gl~866   &                 & 294.3$\pm$3.5 &                 &               \\ \hline 
\end{tabular}
\caption{Trigonometric parallaxes used for the orbital adjustment.
All values are in mas, the references are Van Altena et al. 
(\cite{vanaltena95}, Yale); ESA (\cite{esa97}, HIPPARCOS catalog);
Probst (\cite{probst77}) and Soderhjelm (\cite{soderhjelm99}). The HIPPARCOS
value listed for Gl~644 corresponds to its common proper motion companion 
Gl~643, as the measurement for Gl~644 itself is affected by its 
unaccounted orbital motion.}
\label{photo}
\end{table}

\section{Orbital elements and mass measurements}

\subsection{Orbital adjustment}

All orbits were determined with the ORBIT program 
(Forveille et al. \cite{forveille99}), through a least square adjustment 
to all
available observations: radial velocities or correlation profiles, 
angular separations,
and trigonometric parallaxes. ORBIT supports triple systems, as
well as double ones, as long as three-body effects can be neglected. 
In the present paper this feature was used for two systems, Gl~644 
and Gl866. For double and triple-lined systems we directly adjusted 
the orbit to the cross-correlation profiles (Forveille et al. 
\cite{forveille99}, for 
details), rather than use the radial velocities listed in 
Table~\ref{ind_meas1} to \ref{ind_meas7}. 
This significantly improves the accuracy of the orbital parameters, 
by greatly decreasing the effective number of free parameters of the 
overall adjustment. This gain is particularly important:
\begin{itemize}
\item{} for triple 
systems, whose three correlation peaks blend for many
velocity configurations, 
\item{} for large contrast systems, whose weaker component 
is sometimes only detected with a low signal-to-noise ratio, 
\item{} and
for small amplitude pairs, whose peaks remain blended for most of the orbit.
\end{itemize}
 
\begin{table*}
\tabcolsep 0.85mm
\begin{tabular}{|l|llllllllll|} 
\hline
Name            & \multicolumn{1}{c}{P}             & \multicolumn{1}{c}{T$_0$}
& \multicolumn{1}{c}{e}             & \multicolumn{1}{c}{$\Omega$}    & \multicolumn{1}{c}{$\omega$} &\multicolumn{1}{c}{i} & \multicolumn{1}{c}{as}  & \multicolumn{1}{c}{K$_1$}         & \multicolumn{1}{c}{K$_2$}         & \multicolumn{1}{c|}{V$_0$} \\ 
                & \multicolumn{1}{c}{(days)}        &\multicolumn{1}{c}{(Modified Julian day)} &                           & \multicolumn{1}{c}{(degree)} & \multicolumn{1}{c}{(degree)}& \multicolumn{1}{c}{(degree)}  & \multicolumn{1}{c}{arc sec} &       \multicolumn{1}{c}{(${\rm km\,s^{-1}}$)}       &\multicolumn{1}{c}{(${\rm km\,s^{-1}}$)}&         \multicolumn{1}{c|}{(${\rm km\,s^{-1}}$)}  \\ \hline \hline
Gl~234AB & 5889.0  & 51318.& .371    & 30.7  & 223. & 51.8  & 1.04     & 2.27    & 4.45     & 16.79  \\
         &$\pm$32.&$\pm$13.&$\pm$.004&$\pm$0.5&$\pm$2&$\pm$0.7&$\pm$0.01&$\pm$0.06&$\pm$0.15&$\pm$0.06\\
 & & & & & & & & & & \\
YY~Gem   & 0.8142818    & 50556.5614& .0  & & & 86.5    &0.00135     & 121.67   & 120.97  & 1.97\\
         &$\pm$0.0000003&$\pm$0.0004&Fixed& & &$\pm$0.05&$\pm$0.00005&$\pm$0.43&$\pm$0.44.&$\pm$0.24 \\
 & & & & & & & & & & \\
GJ~2069A & 2.771470     & 50579.1907& .0   & & & 86.7    & 0.0028    & 68.15   & 74.24   & 4.34 \\
         &$\pm$0.0000012&$\pm$0.0002& Fixed& & &$\pm$0.4 &$\pm$0.0003&$\pm$0.03&$\pm$0.05&$\pm$ 0.02 \\
%
%
 & & & & & & & & & & \\
Gl~644A-Bab  &627.0   &53943. &.042    &-10.2   &306.0   &160.3   &.2273     & 5.291    &3.33      &14.947 \\
            &$\pm$0.2&$\pm$3.&$\pm$.001&$\pm$0.2&$\pm$1.5&$\pm$0.1&$\pm$.0004&$\pm$0.015&$\pm$0.02&$\pm$0.007 \\
 & & & & & & & & & & \\
Gl~644Ba-b   &2.965509     & 50919.48 & .0209   & &150.   &  164.18  &     & 16.73    & 18.45   & \\
            &$\pm$0.000006&$\pm$0.03 &$\pm$.0008& &$\pm$3.&$\pm$0.08&     &$\pm$0.02&$\pm$0.03& \\
 & & & & & & & & & & \\
Gl~747AB & 2110.0  & 50432.7 & .274    & 85.1   & -29.1  & 77.3   & .2881    & 6.06     & 6.48    & -47.34 \\
         &$\pm$2.5&$\pm$1.3 &$\pm$.002&$\pm$0.1&$\pm$0.3&$\pm$0.2&$\pm$.0005&$\pm$0.01 &$\pm$0.02&$\pm$0.01 \\
 & & & & & & & & & & \\
Gl~831AB & 703.2  & 50456. & .416    & -35.6  &8.7     & 46.   & .1409    & 5.21    & 9.4    & -55.91 \\
         &$\pm$0.6&$\pm$1. &$\pm$.003&$\pm$0.4&$\pm$0.5&$\pm$1.&$\pm$.0005&$\pm$0.02&$\pm$0.1&$\pm$0.02 \\
 & & & & & & & & & & \\
Gl~866AC-B & 822.6  & 51810.3& .439    & -18.8  &158.7   & 112.6  & .3473    & 5.64    & 10.43   & -50.08 \\
           &$\pm$0.2&$\pm$0.4&$\pm$.001&$\pm$0.1&$\pm$0.3&$\pm$0.1&$\pm$.0005&$\pm$0.04&$\pm$0.04&$\pm$0.01 \\
 & & & & & & & & & & \\
Gl~866A-C  & 3.786516    & 50799.8080& .0   &  & &  116.5   &  & 32.03   & 40.9   & \\
           &$\pm$0.000005&$\pm$0.0005& Fixed&  & &$\pm$0.5&  &$\pm$0.02&$\pm$0.1& \\ \hline
\end{tabular}
\caption{Orbital elements of the newly adjusted orbits. The inclination
angles ($i$) of the two eclipsing binaries, YY~Gem and  GJ~2069, are 
fixed to the values derived from analyses of their light curves (from
Leung \& Schneider \cite{leung78} and Delfosse et al. \cite{delfosse99a}, 
respectively). Their other 
orbital elements are derived from the radial-velocity correlation profiles. 
All orbital elements of the others stars are simultaneously adjusted to the
radial-velocity, parallax, and angular separation data. 
The inner orbits of Gl~644 and Gl~866 have their inclinations $i$ determined 
by requiring that the total mass of the inner binary, derived from 
the outer orbit, must match the sum of the two spectroscopic 
M${\times}\sin^3{i}$ obtained from the inner orbit. 
This leaves an ambiguity between $i$ and $|180-i|$, which we have tentatively
resolved by assuming the approximate coplanarity of their two orbits.
}
\label{orb_el}
\end{table*}

\begin{figure*}
\begin{tabular}{cc}
\multicolumn{2}{c}{Gl 234 (radial-velocity and visual orbit)} \\
\psfig{height=5.7cm,file=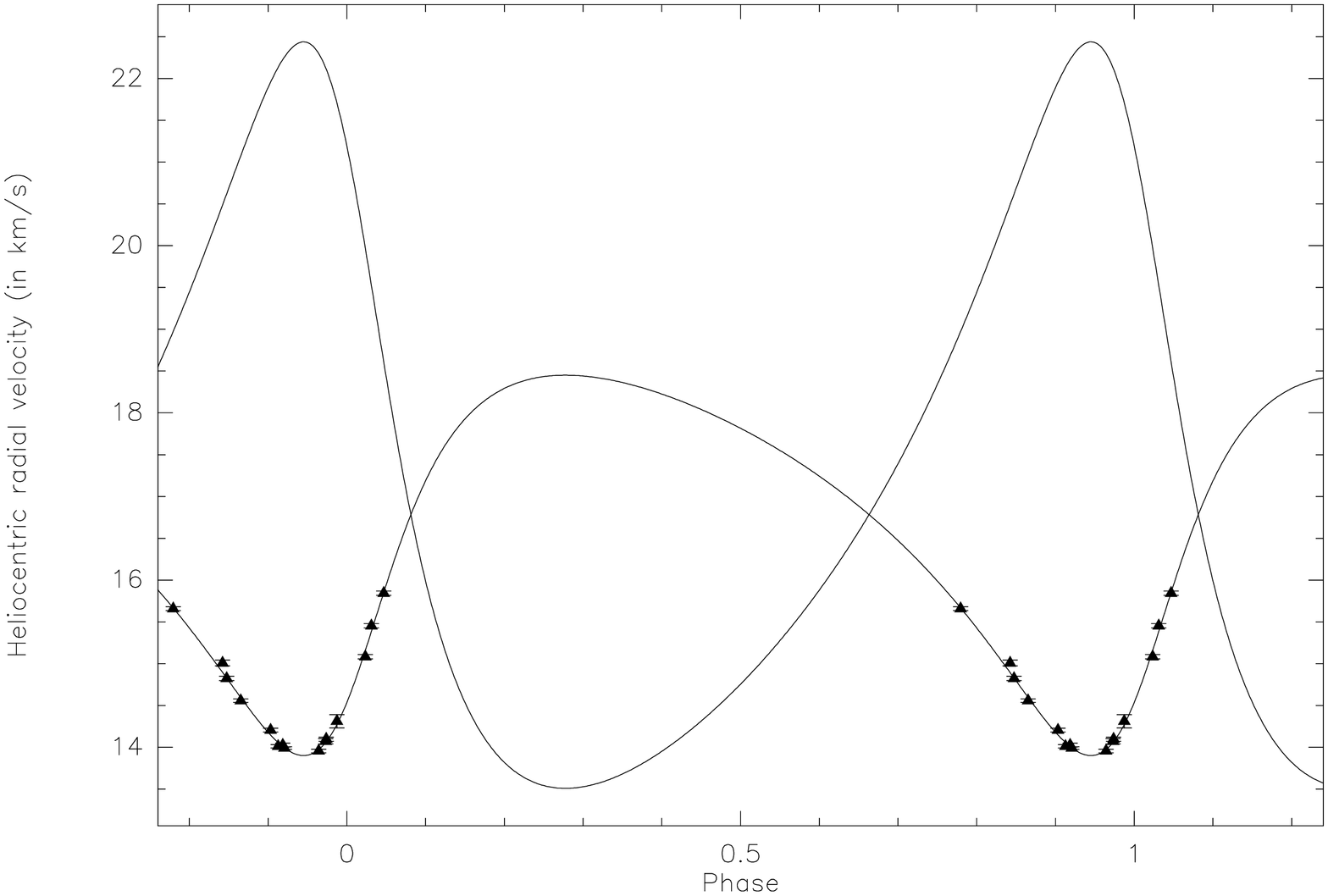,angle=-90} &
\psfig{height=5.7cm,file=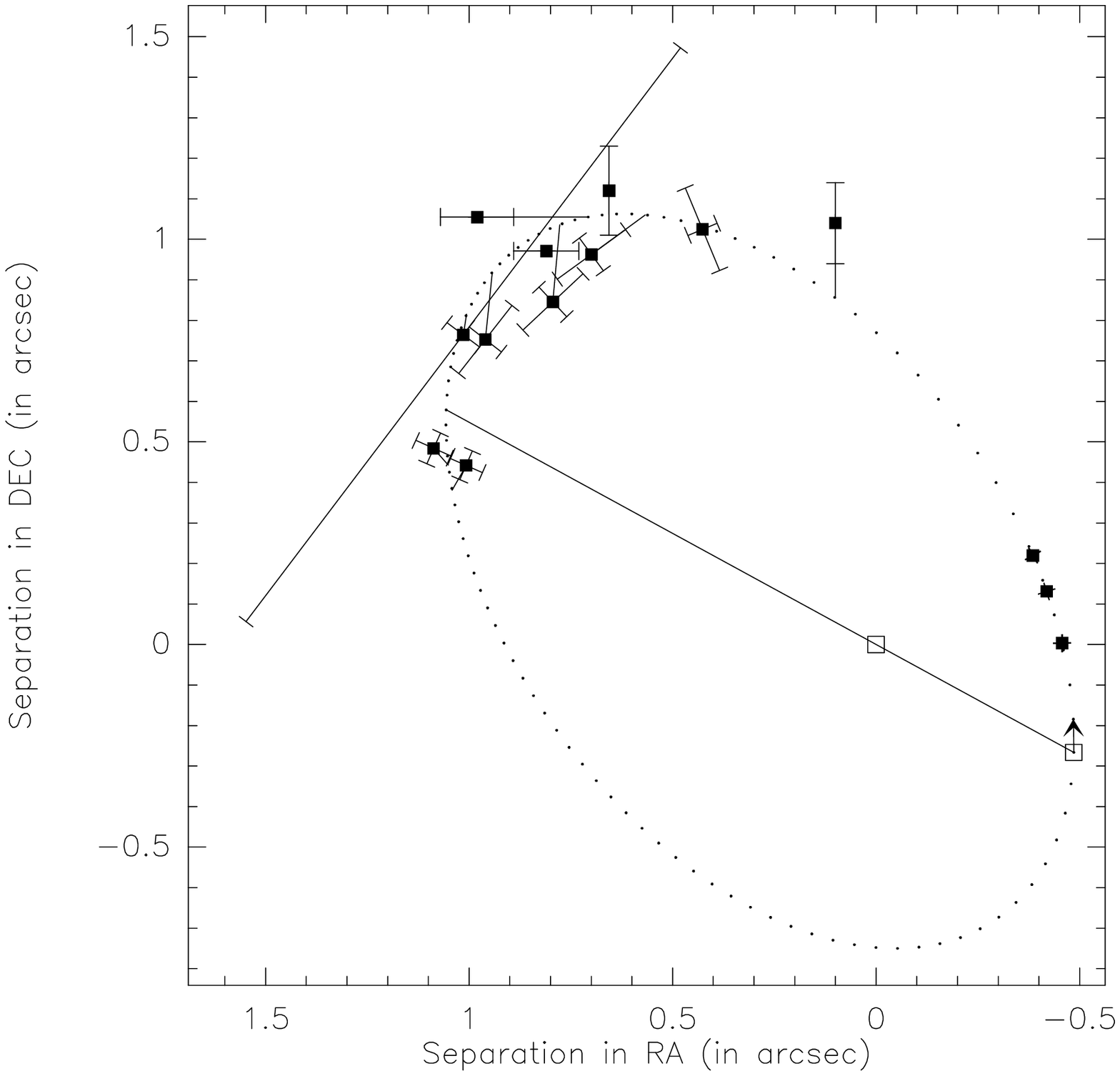,angle=-90} \\
YY Gem (radial-velocity orbit) & GJ 2069A (radial-velocity orbit) \\
\psfig{height=5.7cm,file=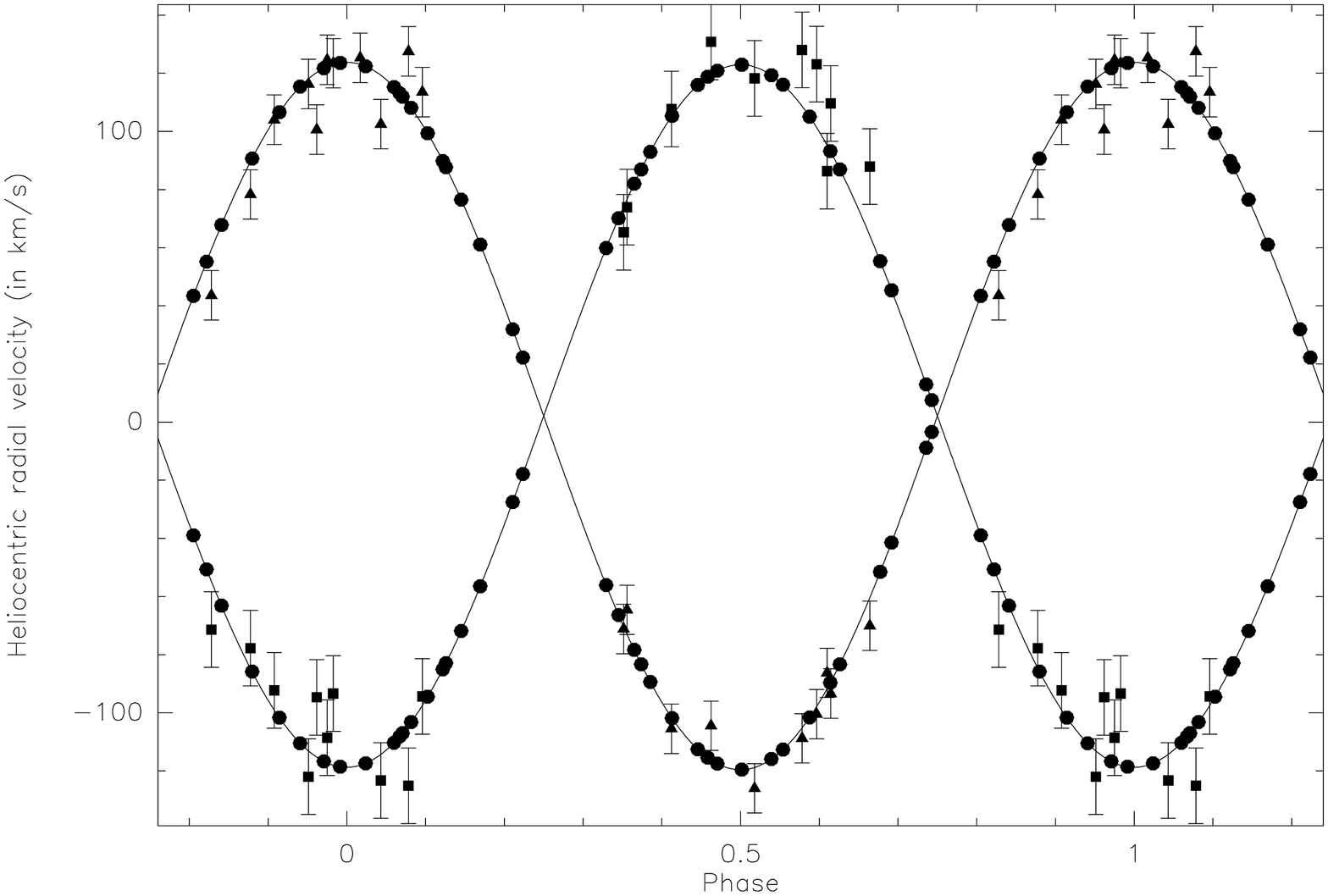,angle=-90} &
\psfig{height=5.7cm,file=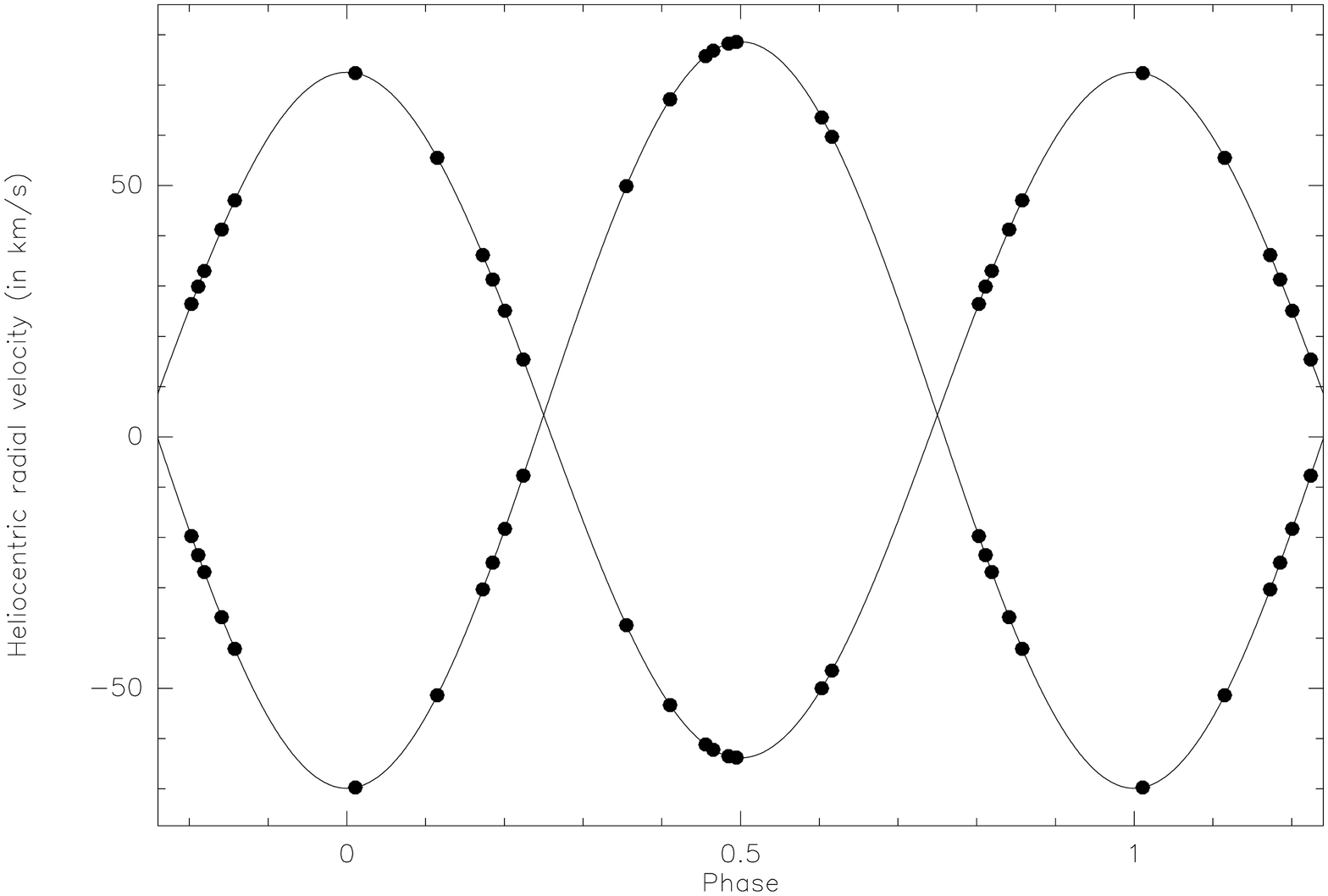,angle=-90} \\
\multicolumn{2}{c}{Gl 747 (radial-velocity and visual orbit)} \\
\psfig{height=5.7cm,file=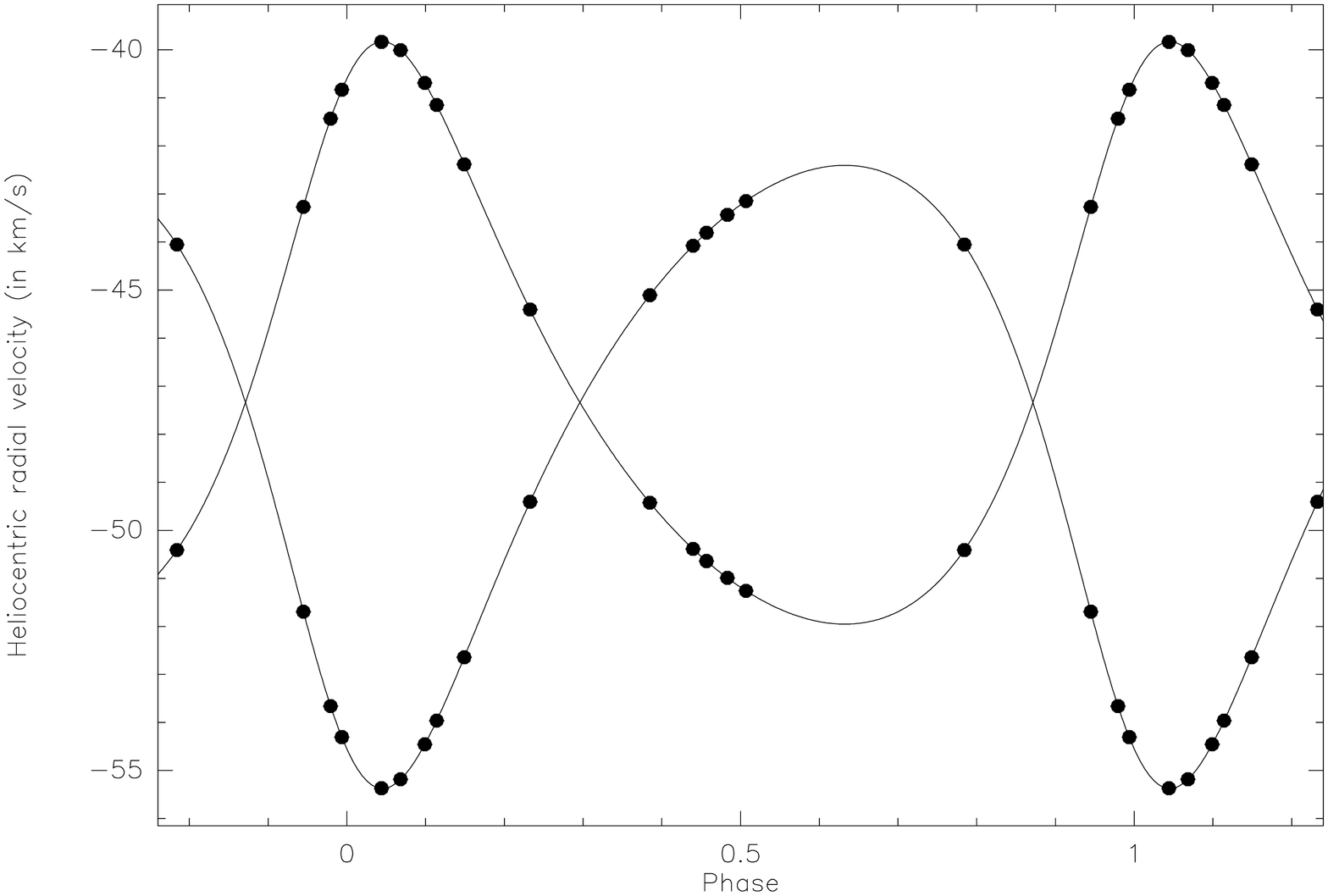,angle=-90} &
\psfig{height=5.7cm,file=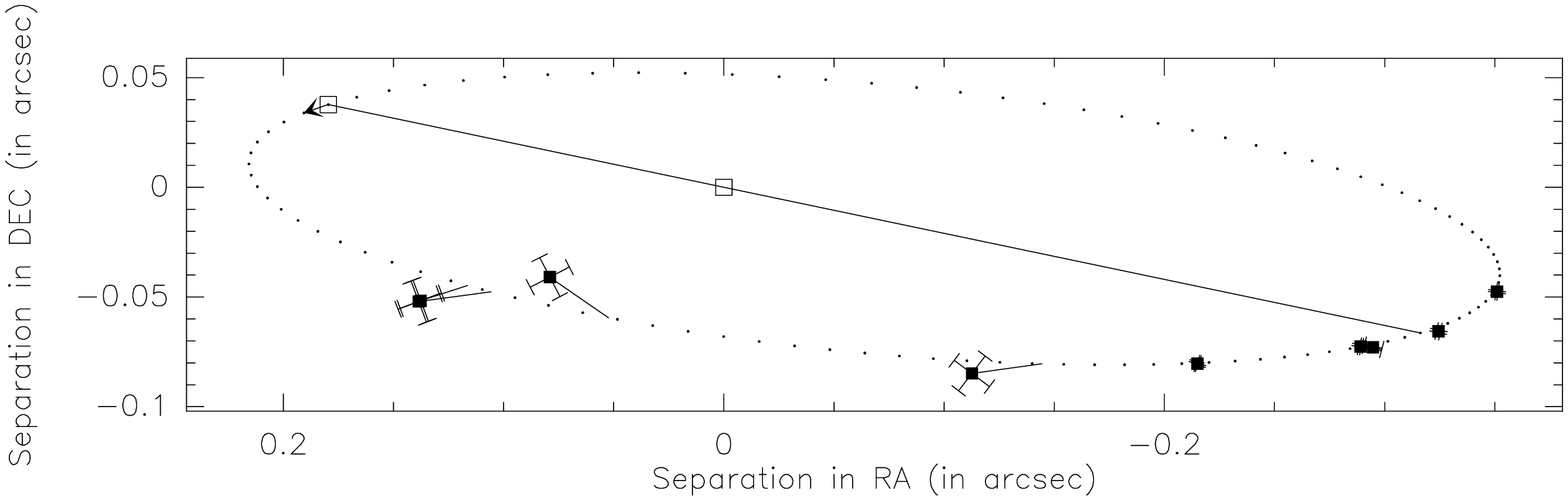,angle=-90} \\
\multicolumn{2}{c}{Gl 831 (radial-velocity and visual orbit)} \\
\psfig{height=5.7cm,file=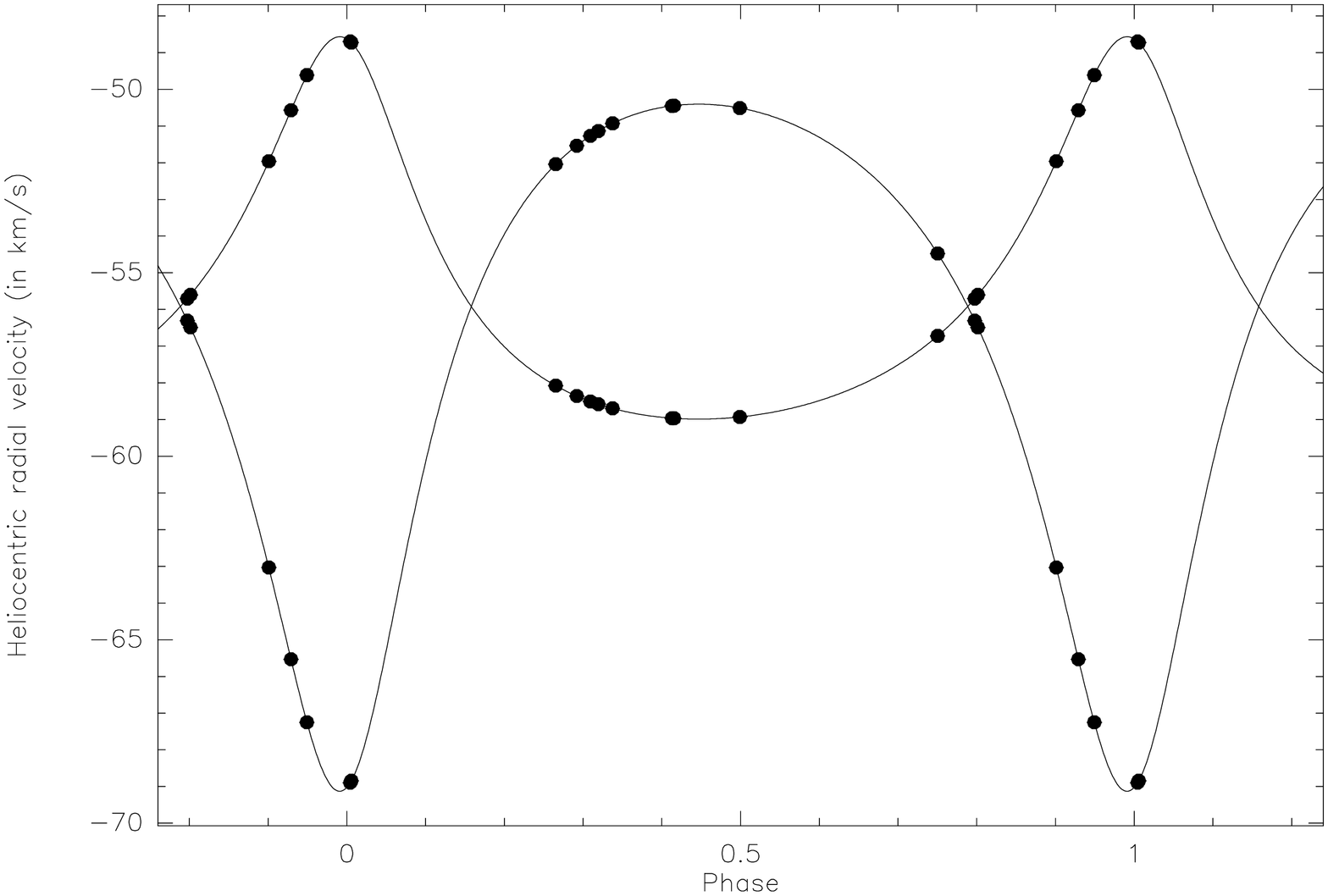,angle=-90} &
\psfig{height=5.7cm,file=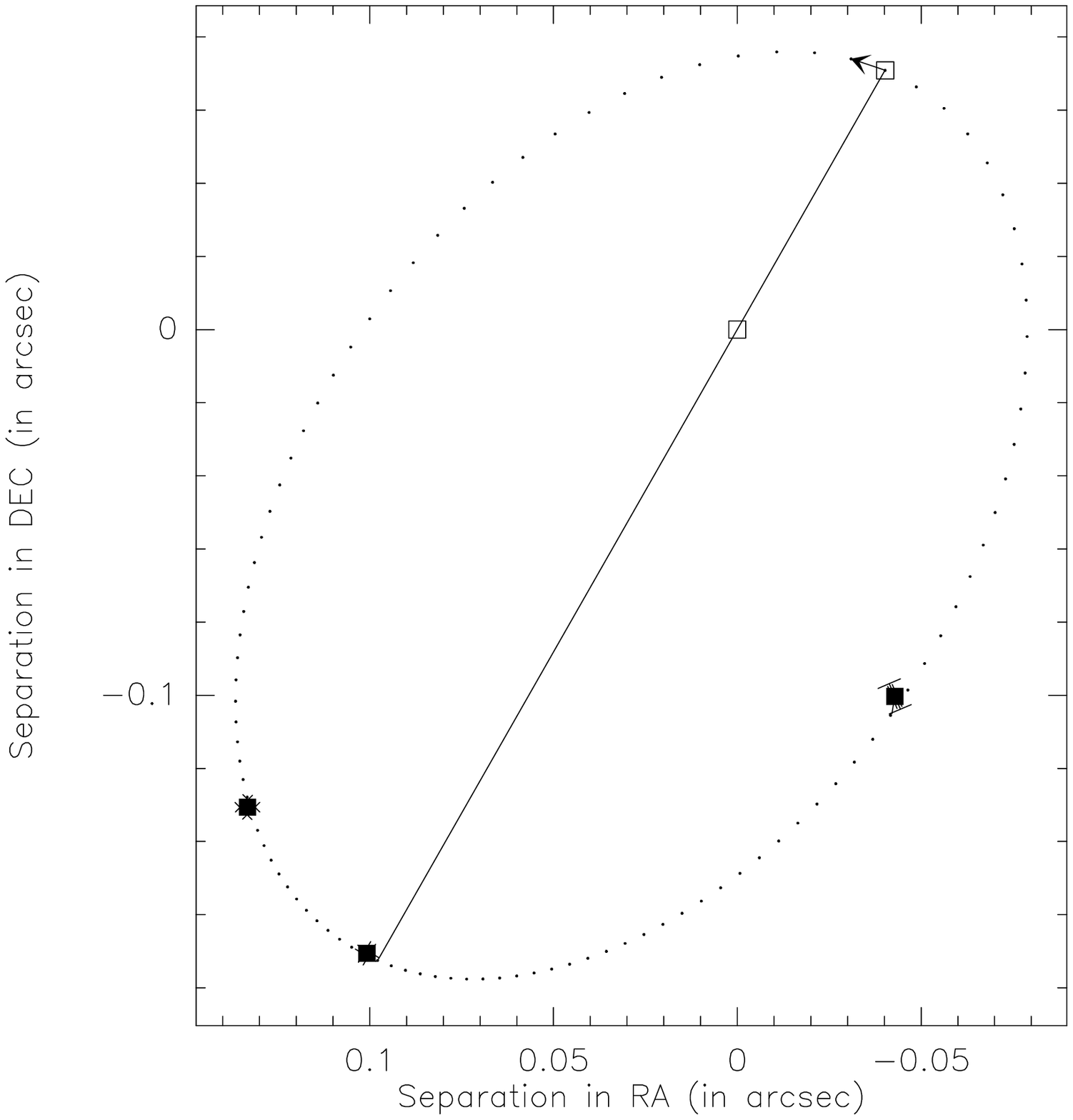,angle=-90} \\
\end{tabular}
\caption{Radial-velocity and visual orbit for the systems with new
or improved mass determinations.}
\label{fig_orb1}
\end{figure*}

\begin{figure*}
\begin{tabular}{cc}
Gl 644 (radial-velocity outer orbit) & Gl 644 (visual orbit) \\
\psfig{height=5.7cm,file=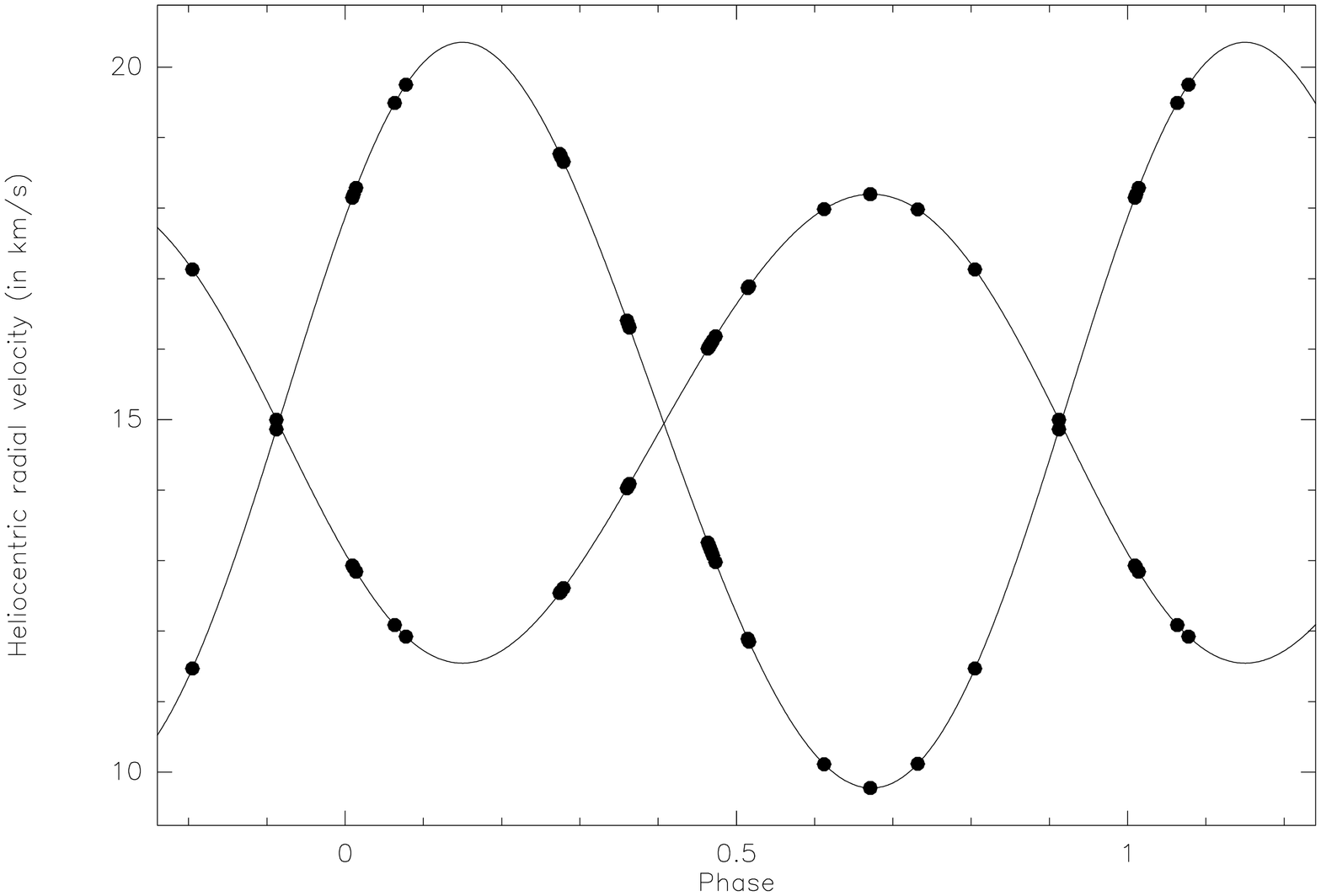,angle=-90} &
\psfig{height=5.7cm,file=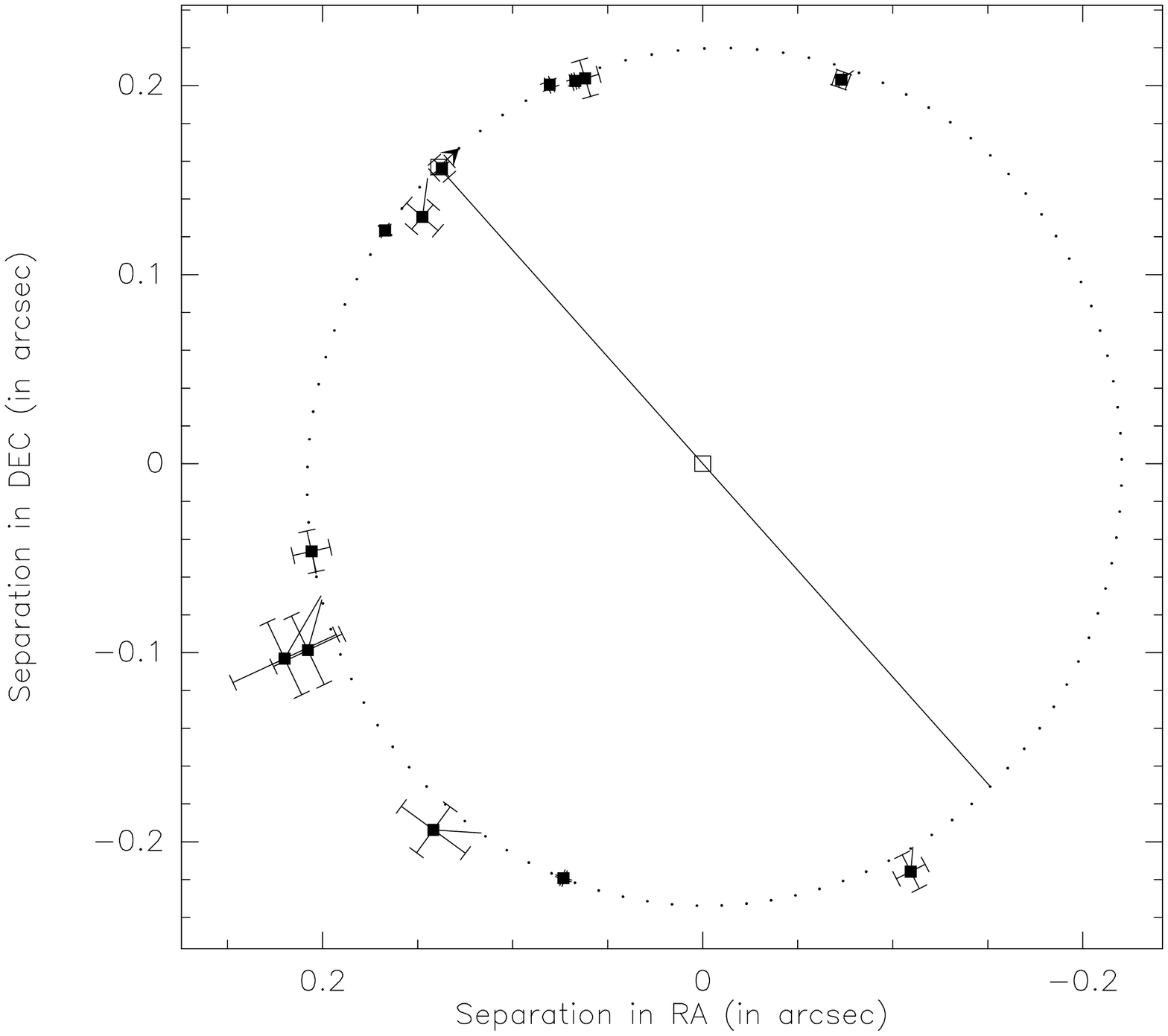,angle=-90} \\
Gl 644 (radial-velocity inner orbit) & Gl 866 (radial-velocity inner orbit) \\
\psfig{height=5.7cm,file=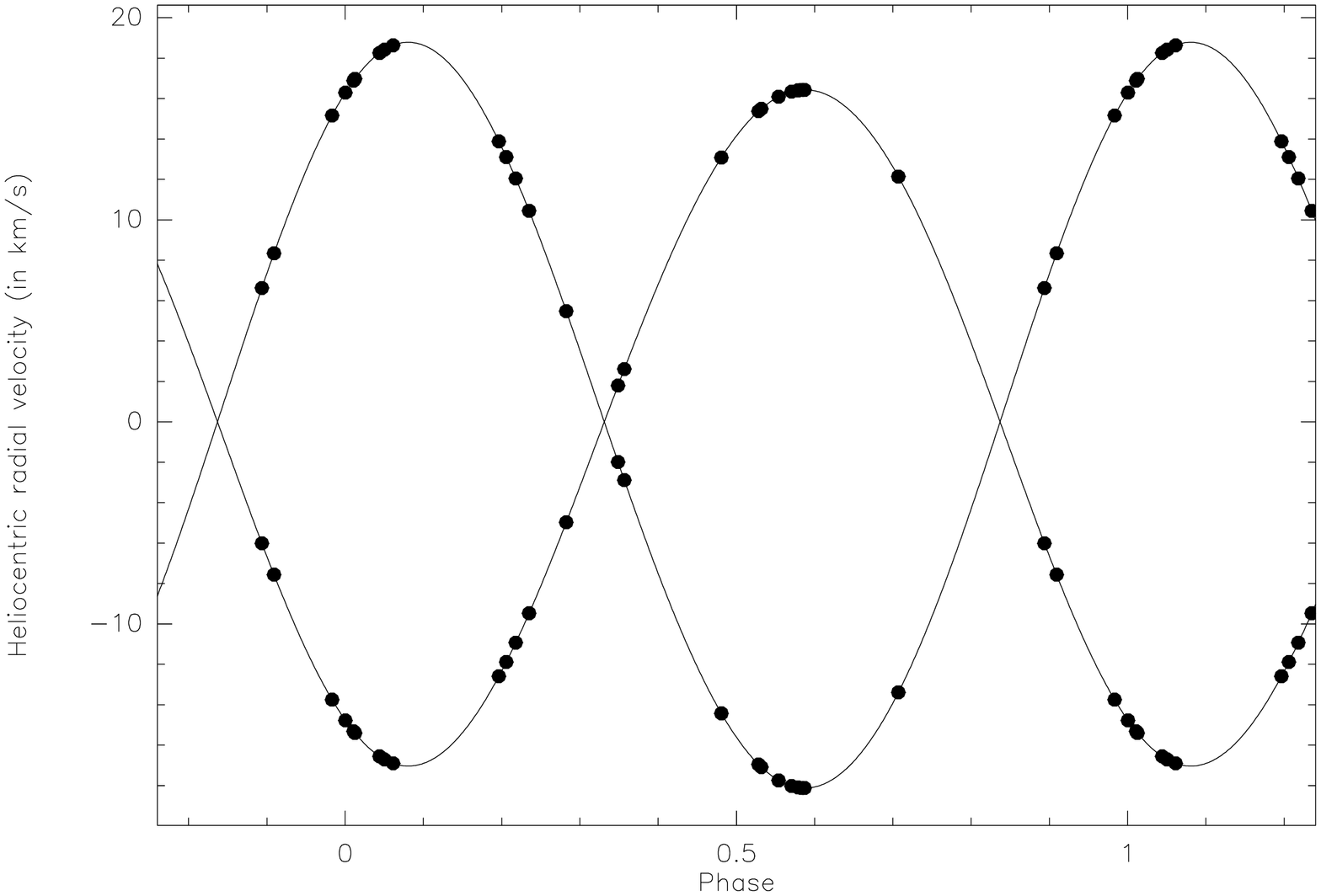,angle=-90} &
\psfig{height=5.7cm,file=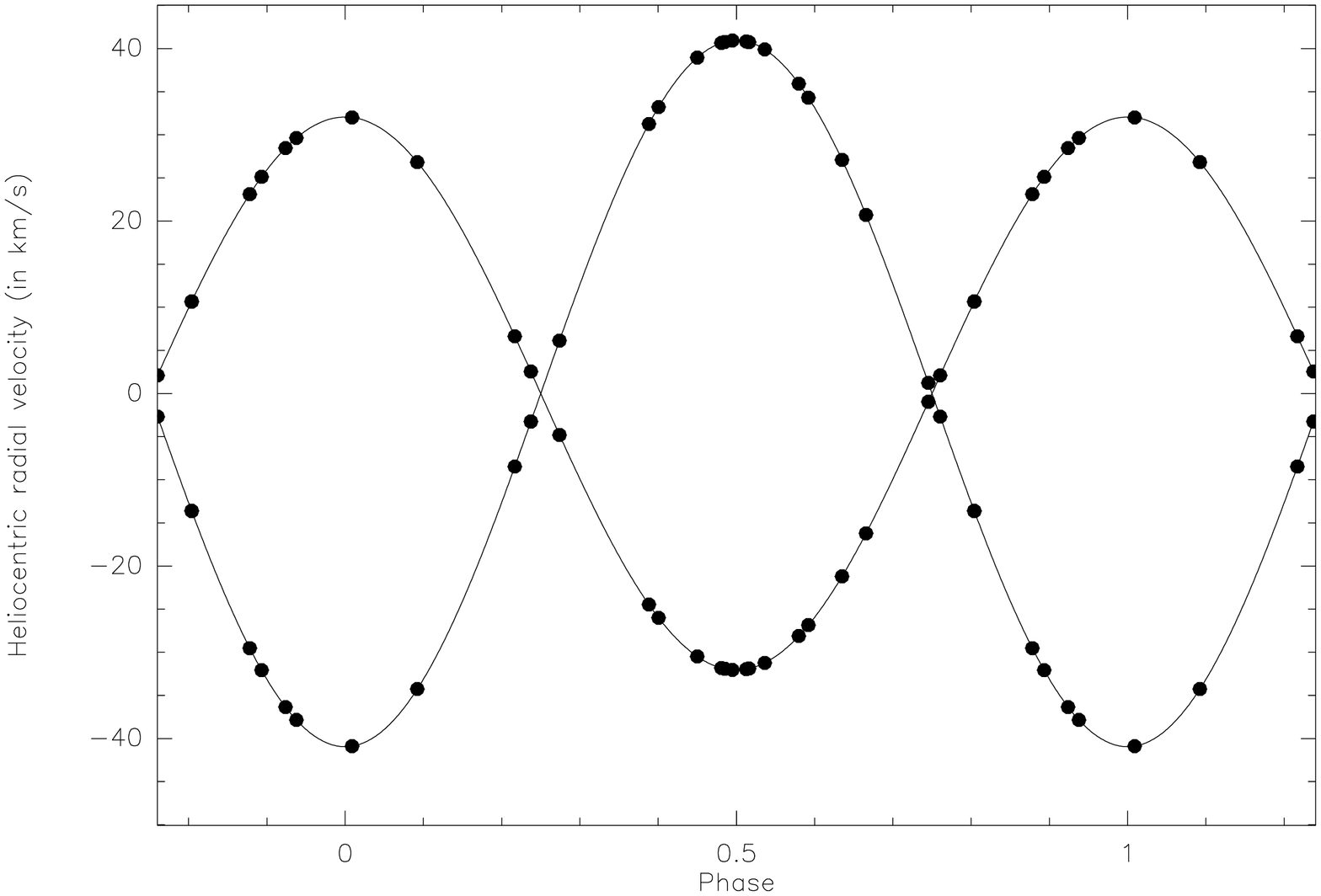,angle=-90} \\
Gl 866 (radial-velocity outer orbit) & Gl 866 (visual orbit) \\
\psfig{height=5.7cm,file=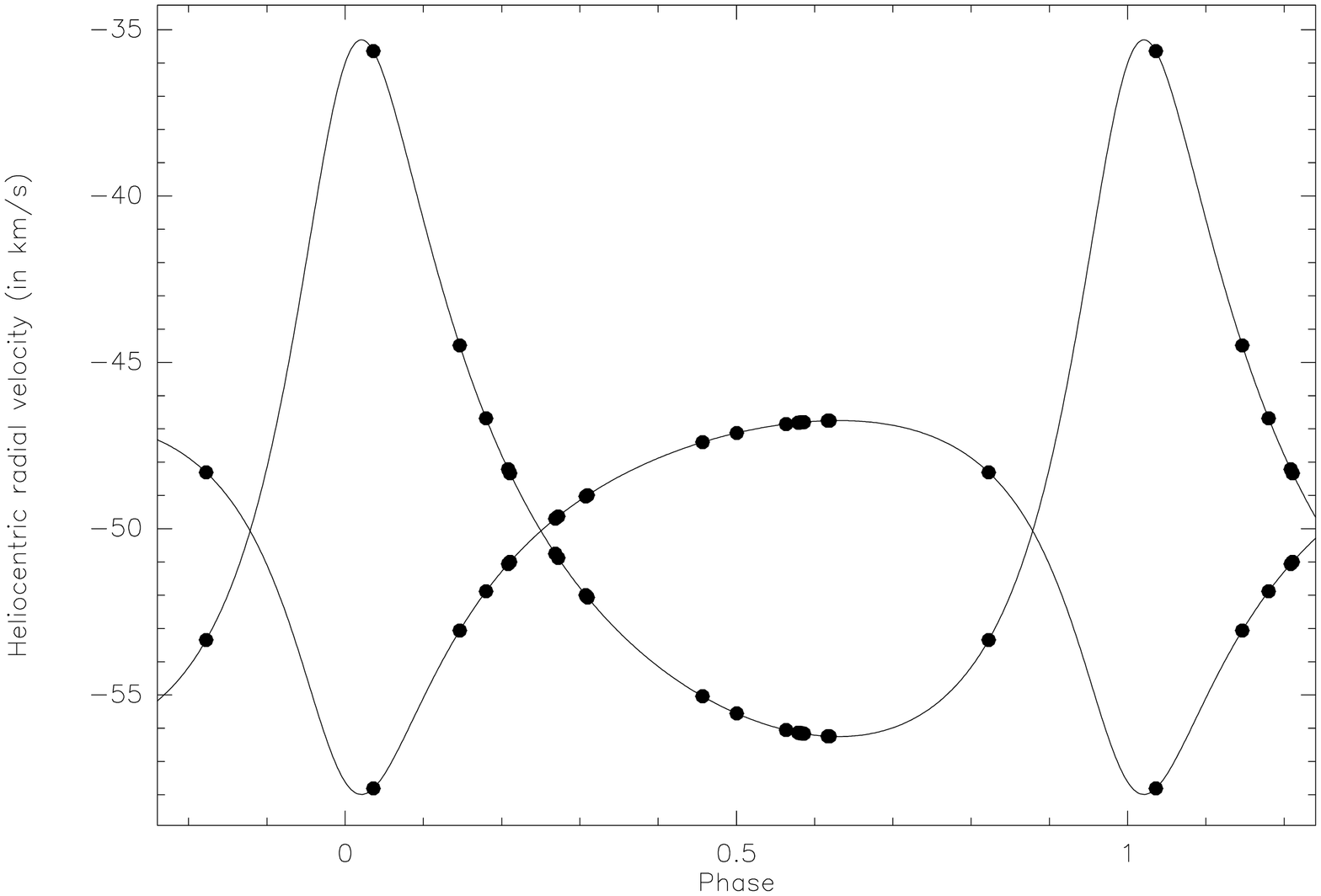,angle=-90} &
\psfig{height=5.7cm,file=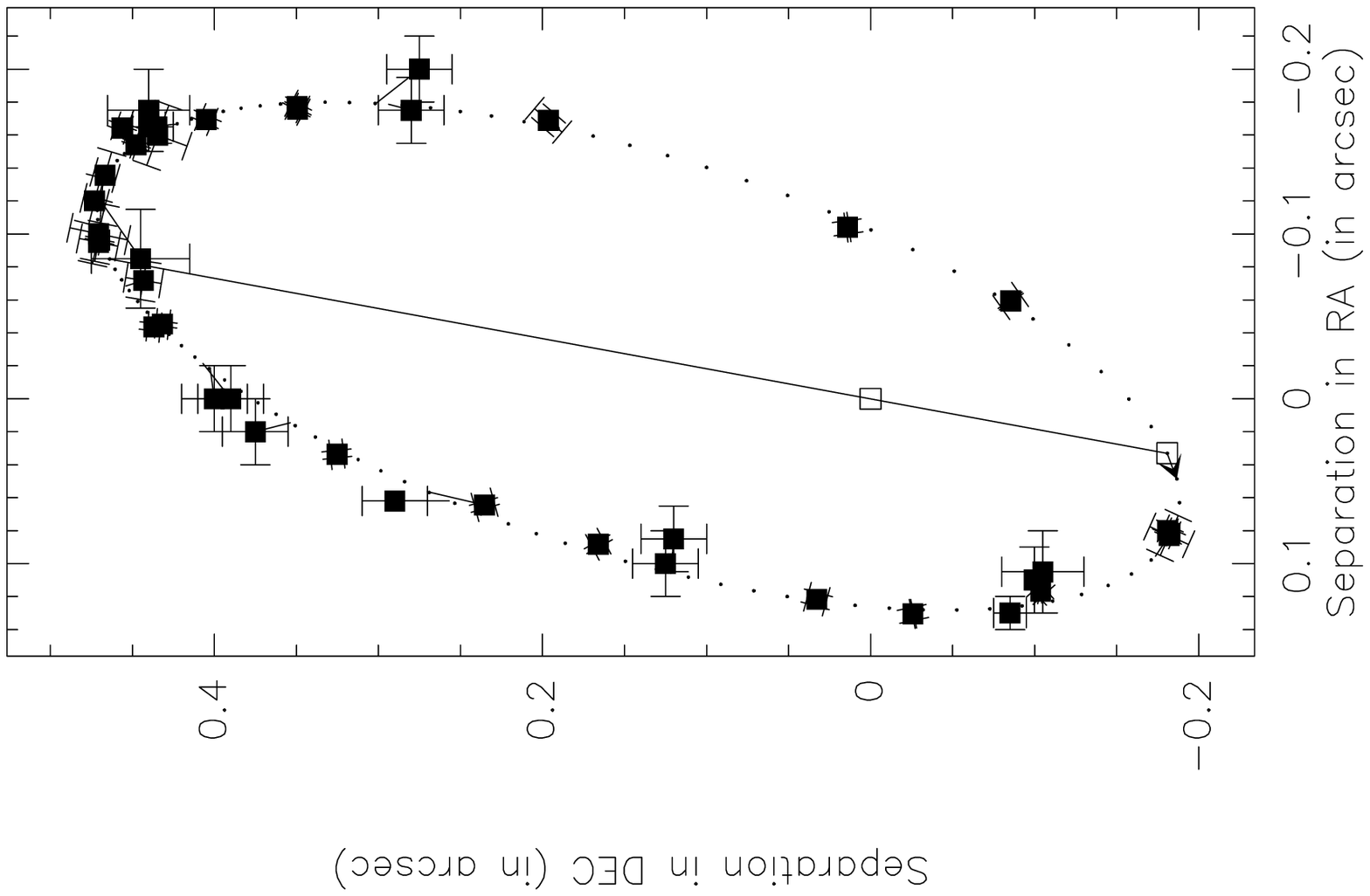,angle=-90} \\
\end{tabular}
\caption{Radial-velocity and visual orbit for the systems with new
or improved mass determinations.}
\label{fig_orb2}
\end{figure*}

Table \ref{orb_el} lists the orbital elements of the 7~systems for which 
we obtained a new or significantly improved orbit. Table~\ref{mass} 
lists the corresponding orbital parallaxes and masses, 
whose relative accuracies range between 
0.2\% and 5\%. 
Fig.~\ref{fig_orb1} and \ref{fig_orb2} respectively show the individual 
radial-velocity curves and visual orbits, which we now briefly discuss.

\subsection{Individual objects}
\label{individual}

\subsubsection{Gl~234 (Ross~614)}

This well known binary is the longest period system (P~=~16.5~years) for 
which we obtain a significantly improved orbit, thanks to the availability
of early measurements that complement the more accurate data we obtained
around the 1999 periastron. The system  was initially 
discovered as an astrometric binary (Reuyl, \cite{reuyl36}), and
intensively studied as such. Probst (\cite{probst77}) 
is usually considered as the current reference astrometric orbit.
Gl~234 was visually resolved on a few occasions (Lippincott \& Hershey,
\cite{lippincott72}), but its 3.5~magnitude contrast made it a difficult
target for visual observers. With the benefit
of hindsight, the masses of 0.11 and 0.06\Msol derived 
by Probst (\cite{probst77}) turn out to have been underestimated 
by a large factor. The system was subsequently resolved in speckle observations
by Mc Alister \& Hartkopf (\cite{mcalister88}) and Coppenbarger et al. 
(\cite{coppenbarger94}). Coppenbarger et al. (\cite{coppenbarger94})
combined these observations with the astrometric orbit
of Probst (\cite{probst77}) to derive masses of M$_A$=0.179$\pm$0.047\Msol 
and M$_B$=0.083$\pm$0.023\Msol, compatible within 1$\sigma$ with the
values listed in Table~\ref{mass}.

Our orbit (Table \ref{orb_el}) is adjusted to the visual data from 
Probst (\cite{probst77}), to the 1D and 2D speckle measurements from 
Coppenbarger et al. (\cite{coppenbarger94}), to our
more accurate adaptive optics
angular separations obtained, to parallaxes 
from Probst (\cite{probst77})
and Soderhjelm (\cite{soderhjelm99}), and to 14 ELODIE radial velocities
of the primary (typical accuracy of 50~m/s). 
The two masses (Table~\ref{mass}) are determined with accuracies of
5.2\% for the primary (0.2027$\pm$.0106 \Msol), 
and 3.4\% for the secondary (0.1034 $\pm$ .0035).

\subsubsection{YY Gem}
YY~Gem is one of the three known detached M-dwarf eclipsing  binaries
(Bopp \cite{bopp74}, Leung \& Schneider \cite{leung78}). 
We have obtained 75 radial-velocity measurements of the two components 
with ELODIE, with typical 
standard errors of 2~km/s. Both components of YY~Gem have their rotation
period synchronized with the short orbital period by tidal interactions.  
The resulting fast equatorial velocities (v~sin~{\it{i}}$\sim$30~km/s)
explains this very degraded velocity accuracy. The noisier measurements of
Bopp (\cite{bopp74}) ($\sim$~10~km/s) are also used in the adjustment
and help to constrain the period. The amplitudes, $K1$ and $K2$,
on the other hand are almost completely determined by the ELODIE 
measurements, and so are therefore the masses, determined with 
relative accuracies of 0.2\% . We are now 
obtaining infrared lightcurves of this system to improve the 
determination of the two stellar radii, and will then present a complete 
reanalysis.

\begin{table}
\tabcolsep 1.6mm
\begin{tabular}{|ll|ll|} \hline
\multicolumn{2}{|l|}{Name} & Mass (in \Msol) & parallax (in mas) \\ \hline
Gl 234 & A & .2027 $\pm$ .0106 (5.2 \%) & 243.7 $\pm$ 2.0 (0.8\%)\\
       & B & .1034 $\pm$ .0035 (3.4\%) &  \\
YY Gem & a & .6028 $\pm$ .0014 (0.2\%) &  \\ 
       & b & .6069 $\pm$ .0014 (0.2\%) &  \\
GJ 2069A& a & .4344 $\pm$ .0008 (0.2\%) &  \\
       & b & .3987 $\pm$ .0007 (0.2\%) &  \\
Gl 644 & A & .4155 $\pm$ .0057 (1.4\%) & 154.8 $\pm$ 0.6 (0.4\%) \\
       & Ba & .3466 $\pm$ .0047 (1.3\%) &  \\
       & Bb & .3143 $\pm$ .0040 (1.3\%) &  \\
Gl 747 & A & .2137 $\pm$ .0009 (0.4\%) & 120.2 $\pm$ 0.2 (0.2\%) \\ 
       & B & .1997 $\pm$ .0008 (0.4\%) &     \\
Gl 831 & A & .2913 $\pm$ .0125 (4.3\%) & 117.5 $\pm$ 2.0 (1.7\%) \\  
       & B & .1621 $\pm$ .0065 (4.0\%) &  \\
Gl 866 & A & .1187 $\pm$ .0011 (0.9\%) &  293.6 $\pm$ 0.9 (0.3\%) \\
       & B & .1145 $\pm$ .0012 (1.0\%) &  \\
       & C & .0930 $\pm$ .0008 (0.9\%) &  \\ \hline
\end{tabular} 
\caption{Masses and parallaxes derived from the orbits listed in 
Table~\ref{orb_el}. Except for the two eclipsing systems,
the parallaxes represent an optimally weighted combination of the astrometric 
and orbital parallaxes. Some of them, such as Gl~644, merely reflect the 
astrometric parallax.
Others, such as Gl~866, are almost completely determined by the orbit.
}
\label{mass}
\end{table}

\subsubsection{GJ~2069A (CU Cnc)}

GJ~2069A is one of the three known detached M-dwarf eclipsing binaries
(Delfosse et al. \cite{delfosse99a}). We present here an improved orbit, 
which includes a few radial-velocity measurements
obtained after the completion of Delfosse et al. (\cite{delfosse99a}). More
importantly, the new orbit was directly adjusted to the ELODIE 
cross-correlation profiles, whereas our earlier article adjusted an orbit 
to radial velocities extracted from these profiles. The resulting
masses are now among of the most accurate measured for any star 
(e.g. Andersen \cite{andersen91}, \cite{andersen98}), with 0.2\% accuracies 
for both components. It is somewhat unfortunate
that they are not yet matched with equally precise distance, infrared 
photometry, metallicity, and radii.

We have recently discovered (Beuzit et al., in prep.) a fainter 
companion to the GJ~2069Aab pair, at a separation of 0.55" in early 2000 and
which we name GJ~2069D. This makes GJ~2069 a quintuple system, since we had 
earlier found the fainter visual component, GJ~2069B, to be an adaptive 
optics and spectroscopic binary (Delfosse et al. \cite{delfosse99c}). 
The new companion is 3 magnitudes fainter than GJ~2069Aab in the K band. 
Its influence on the 
photometry can therefore safely be neglected at the current precision 
of the absolute magnitudes, and the extrapolated luminosity contrast in 
the V band precludes its detection in the integrated visible spectrum. 
GJ~2069D will also eventually cause a drift in the 
systemic velocity of GJ~2069Aab. We have attempted to fit this drift
in addition to the parameters of the Aab orbit, but we found that this does not
decrease the ${\chi}^2$ of the adjusted model. This indeterminacy of the
drift parameter indicates that the period of the AD system is significantly
longer than the $\sim$4~years span of the current radial-velocity data. Its 
influence  does therefore not appreciably bias the measured masses.

\subsubsection{Gl~644}

At d~=~6.5pc, the Gl~644/643 system is the richest stellar system in the 
immediate solar neighbourhood (d$<$10pc), with 5 components 
Gl~644 (M3V), is the brightest visual component and
shares a common proper motion with two distant companions, Gl~643 (M3.5V) 
at 72 arc seconds and vB8 (Gl~644C, M7V) at 220 arc seconds. Gl~644 is 
itself a 1.7-year binary, identified from astrometric observations by 
Weiss (\cite{weiss82}) and Heintz (\cite{heintz84}), and first 
angularly resolved in 
speckle observations by Blazit et al. (\cite{blazit87}) 
and Tokovinin \& Ismailov (\cite{toko88}). Finally, 
Pettersen et al. (\cite{pettersen84}) found that one of the two 
components of Gl~644 (Gl644B) is 
itself a short period spectroscopic binary, but could not determine 
its period.

We have obtained 25 ELODIE radial-velocity measurements of Gl~644, which
usually appears as a well separated triple-lined system in those data.
This allows us to determine for the first time the elements of the inner 
orbit, whose period is 2.97 days. Such close orbits are very rapidly
circularised by tidal interactions. Here we nonetheless measure a small but 
highly significant eccentricity of .0209$\pm$.0008. This most likely
results from dynamical interactions between the two orbits (Mazeh \& 
Shaham \cite{mazeh79}), as could
probably be ascertained through a complete dynamical analysis (which 
would be beyond the scope of the present paper). 

The orbital elements listed in Table~\ref{orb_el} were simultaneously
determined for the two orbits, using angular separation measuremements 
from Blazit et al. (\cite{blazit87}), Tokovinin \& Ismailov 
(\cite{toko88}), Al-shukri et al. (\cite{alshukri96}), Balega et
al. (\cite{balega89}, \cite{balega91}, \cite{balega94}), Hartkopf et al. 
(\cite{hartkopf94}), our own more accurate adaptive optics measurements 
(Table~\ref{ind_meas9}), the 25 radial-velocity profiles, and the 
trigonometric
parallaxes of both Gl~644 (Soderhjelm \cite{soderhjelm99})
and Gl~643 (ESA \cite{esa97}). This determines the masses of all three 
components with relative accuracies of 1.4 - 1.3\%. 

This accuracy is obtained even though the orbit is almost seen face-on 
($i=160.3^{\circ}$), an adverse orientation
for accurate mass measurements from radial velocities. 
ELODIE generally provides very accurate $M.\sin^3(i)$ for 
double-lined systems, and the mass errors are then typically
dominated by the inclination uncertainty.
For nearly edge-on orbits, errors on the inclination do not propagate
much to $\sin (i)$, and we can then obtain accurate masses 
even for fairly uncertain inclinations. For nearly face-on orbits 
on the opposite, one needs a very accurate inclination to determine 
even moderately accurate masses. Gl~644 demonstrates that we can
obtain accurate masses even in some rather poorly oriented orbits.

Interferometric measurements would be needed to resolve the very close 
inner orbit, but its inclination $i_s$ is nonetheless strongly constrained: 
M$_{\rm{Ba}}+$M$_{\rm{Bb}}$, as derived from the outer orbit, must
match the sum of the two spectroscopic M${\times}sin^3{i_s}$ obtained from 
the inner orbit. This gives $\sin (i_s)$=0.27, and therefore either
$i_s$=164.2$^o$ or $i_s$=15.8$^o$. One of those two determinations
is very close to the inclination of the outer orbit ($i_s-i_o~{\sim}~4^o$).
This probably points to a coplanar system, in keeping with a general 
tendency of close triple systems (Fekel \cite{fekel81}). To ascertain this, 
one would need to resolve the inner pair, determining ${\Omega}_s$ and 
obtaining $i_s$ without the reflexion ambiguity.

\subsubsection{Gl~747AB (Kui~90)}
Gl~747 was first visually resolved in 1936 by Kuiper. Yet, the orbit
of this nearby star (d~=~8.5~pc) has apparently never been determined,
probably because its separation never exceeds 0.35''. It has been resolved
in speckle observations once by each of Blazit et al. 
(\cite{blazit87}) and Mc Alister et al. (\cite{mcalister87}), and 
three times by Balega et al. (\cite{balega89}). 
We have complemented
these litterature measurements with 15 ELODIE radial-velocity profiles 
of this  double-lined system, and 4 separations obtained with PUE'O.
The 5.5-year period orbit listed in Table~\ref{orb_el} provides
an excellent description of all these measurements, with the (strong) 
exception of the speckle separation obtained by Mc Alister et al. 
(\cite{mcalister87}). We could not identify a likely reason for this 
discrepancy, except that Gl~747 is significantly fainter than most 
sources in Mc Alister et al. (\cite{mcalister87}). It could 
have been close to their sensitivity limit for the conditions under which
it was observed, but is on the other hand a system of two equally bright
stars. It should therefore not have been an overly difficult target for
speckle observations. We have chosen to ignore this data point, since 
all other measurements are mutually consistent to within approximately 
their stated standard errors, and since some of them have been observed 
within a year of the discrepant point.
This orbit determines the masses of the two components (2$\times$0.2\Msol)
with an accuracy of $\sim$0.4\%, and the orbital parallax with a 0.2~mas
standard error. The latter is in excellent agreement 
with the less precise astrometric parallax listed in Van Altena et al.
(\cite{vanaltena95}).

\subsubsection{Gl~831}
Gl~831 was first noticed as a P=1.93-year astrometric binary 
(Lippincott \cite{lippincott79}, Mc Namara et al. 
\cite{mcnamara87}), and then resolved by 
visible speckle observations (Blazit et al. \cite{blazit87}). 
It appears in ELODIE observations as a double-lined spectroscopic
binary, but the contrast between the two peaks of the correlation 
function is large ($\sim$10) and most of the time their separation
is not very much larger than their combined width. It is therefore
a good illustration of the improvement brought by a direct adjustment
of the orbit to the correlation profiles. Using three adaptive optics
angular separations, %
the parallax from  Van Altena et al. (\cite{vanaltena95}) 
and 14 ELODIE correlation profiles we determine
both masses with 4\% relative accuracy.

Henry et al. (\cite{henry99}) have found tentative evidence for a third 
component of Gl~831 in their {\it HST} FGS observations of this system.
This companion would be $\sim$3 magnitude fainter in the V band than 
the primary, and it could be either very close to A or B, or beyond
$\sim$0.5''.
We can now firmly exclude the first possibility for a physical 
member of Gl~831: it would necessarily imply very large velocity variations 
of the corresponding bright component. A red companion that is only three 
magnitude fainter in the V band than the primary would be easily
detected in our K band adaptive optics images, unless it was
always fortuitously within $\sim$0.15'' of the primary or within 0.1''
of Gl~831B, whenever we observed it. As this is rather unlikely, the 
companion, if real, is most likely bluer than Gl~831. It could then either 
be a white dwarf member of the system, or an unrelated background
object.

\subsubsection{Gl~866}

We recently (Delfosse et al. \cite{delfosse99b}) discussed 
in detail this system 
of three very low mass stars (3~$\times$~$\sim$0.1\Msol), 
and obtained individual masses with $\sim$3\% accuracy. Shortly 
thereafter Woitas et al. (\cite{woitas00}) published a large set 
of new angular 
separation measurements. Lacking radial-velocity information, they could 
only obtain the total mass of the system, with $\sim$10\% accuracy. 
We analyse here the combination of the two datasets, and obtain a very 
substantial improvement over either of these previous analyses. All three
masses are now 
determined with $\sim$1\% accuracy. Gl~866C, the faintest member of 
the system, is the only star with a dynamically determined mass 
(0.0930$\pm$0.0008\Msol) that is safely lower than 0.1\Msol.

\section{Conclusions}

We have presented here mass measurements for 16 VLMS, with 
accuracies which range between 0.5 and 5\%, and for masses 
down to below 0.1\Msol. We will shortly publish a few additional 
masses whose derivation involve long baseline interferometric
measurements 
(S\'egransan et al., in prep.). These results only 
represent a snapshot of  the progress of our mass measurement 
program: we continue 
to monitor the VLMS binaries with high accuracy radial-velocity 
observations and adaptive optics imaging, and have started using
long baseline interferometry. Most of the masses which we present
here will be improved in the future, and additional ones with similar 
accuracies will become available. 

In a companion paper (Delfosse et al. \cite{delfosse00a}), 
we show that these masses provide an impressive validation of the 
theoretical infrared M-L relations
of Baraffe et al. (\cite{baraffe98}), but point towards low level 
($\sim$0.5~mag) 
deficiencies of these models in the V band.

A logical next step for this type of work is the determination of 
accurate masses for even fainter objects, the very late M dwarfs
and the L dwarfs. There is at present a large effort in determining
the mass function across the stellar/substellar boundary in young open
clusters (Bouvier et al. \cite{bouvier98}, 
Zapatero Osorio et al., in prep.) and in the
field (Reid et al. \cite{reid99}, Delfosse \& Forveille \cite{delfosse00}). 
These programs need
an accurate calibration of the mass as a function of both 
luminosity and age (due to the dominant effect of cooling 
for brown dwarfs). 
Up to now such relations are only available from models, which 
unfortunately meet with new difficulties for temperatures lower than 
were relevant in this paper, as dust condenses in the atmospheres of 
very cool dwarfs. 


Until very recently, no binary of such mass was known with 
a period short enough for a mass determination over any realistic
time scale. A few are now known (Mart\'{\i}n et al. 
\cite{martin99}, \cite{martin00},
Koerner et al. \cite{koerner99}), even though their periods are either shorter
or longer than would be ideal for a quick and accurate mass measurement.
They will eventually provide mass determinations for brown dwarfs,
but additional efforts to find more brown dwarfs binaries,
and better suited ones, are certainly more than warranted. Several groups are
doing this, following up the late-M and L dwarfs discovered
by the SLOAN, 2MASS and DENIS surveys.

\begin{acknowledgements}

We thank the technical staff and telescope operators of OHP and CFHT
for their efficient support during these long-term observations. 
This research has made use of the Simbad database, operated at CDS, 
Strasbourg, France, of the WDS catalog maintained at the USNO,
Washington, USA, and of the Third Catalog of Interferometric 
Measurements of Binary Stars. 

\end{acknowledgements}

\begin{table*}
\begin{tabular}{|l|l|l|} 
\multicolumn{3}{c}{Gl234AB}\\
\hline        
Julian Day   &  V$_A$               &Reference \\
2400000+                   & (in km/s)            &\\ \hline \hline
     &    & this paper \\                                       
50018.6620               & 15.658  $\pm$   .050 &\\
50388.6449               & 15.007  $\pm$   .050 &\\
50418.5618                  & 14.824  $\pm$   .050 &\\
50524.3498                  & 14.557  $\pm$   .050 &\\
50747.6659                  & 14.204  $\pm$   .050 &\\
50803.5423                  & 14.010  $\pm$   .050 &\\
50838.4212                  & 14.025  $\pm$   .050 &\\
50851.3844                  & 13.989  $\pm$   .050 &\\
51107.6827                 & 13.955  $\pm$   .050 &\\
51159.5379                 & 14.071  $\pm$   .050 &\\
51163.6391                 & 14.096  $\pm$   .050 &\\
51243.3511                 & 14.310  $\pm$   .080 &\\
51455.6608                 & 15.083  $\pm$   .050 &\\
51500.5519                 & 15.452  $\pm$   .050 &\\
51509.6176                 & 15.322  $\pm$   .050 &\\
51594.3558                 & 15.843  $\pm$   .050 &\\ \hline
\end{tabular}                                                                             
\caption{Radial-velocity measurement of Gl~234A. 
To be published in electronic form only.}
\label{ind_meas1}
\end{table*}

\begin{table*}
\begin{tabular}{|l|l|l|l|} 
\multicolumn{4}{c}{YY Gem}\\\hline         
Julian Day   &  V$_A$               &     V$_B$                &Reference  \\
2400000+     & (in km/s)            & (in km/s)                & \\
\hline\hline                                                                 
             &                      &                          & Bopp 1974\\
40988.819000 &  121.800000 $\pm$ 8.500000 & -86.100000  $\pm$  13.00000&  \\  
40991.858000 &  51.900000  $\pm$ 8.500000 & -63.100000  $\pm$  13.00000&  \\
40992.792000 &  132.900000 $\pm$ 8.500000 & -100.300000 $\pm$  13.00000&  \\
40992.876000 &  135.800000 $\pm$ 8.500000 & -116.800000 $\pm$  13.00000&  \\
40995.741000 &  -92.200000 $\pm$ 8.500000 & 131.400000  $\pm$  13.00000&  \\
40995.796000 &  -61.800000 $\pm$ 8.500000 & 96.200000   $\pm$  13.00000&  \\
40996.784000 &  86.600000  $\pm$ 8.500000 & -69.500000  $\pm$  13.00000&  \\
40996.844000 &  124.600000 $\pm$ 8.500000 & -113.700000 $\pm$  13.00000&  \\
40998.802000 &  -56.300000 $\pm$ 8.500000 & 82.200000   $\pm$  13.00000&  \\
40998.848000 &  -97.200000 $\pm$ 8.500000 & 116.000000  $\pm$  13.00000&  \\
40999.827000 &  -85.100000 $\pm$ 8.500000 & 117.900000  $\pm$  13.00000&  \\
41285.879000 &  112.300000 $\pm$ 8.500000 & -84.000000  $\pm$  13.00000&  \\
41285.923000 &  108.900000 $\pm$ 8.500000 & -86.400000  $\pm$  13.00000&  \\
41285.968000 &  133.600000 $\pm$ 8.500000 &                            &  \\
41292.845000 &  -96.200000 $\pm$ 8.500000 & 139.100000  $\pm$  13.00000&  \\
41292.890000 &  -117.700000$\pm$ 8.500000 & 126.500000  $\pm$  13.00000&  \\
41292.939000 &  -100.500000$\pm$ 8.500000 & 136.300000  $\pm$  13.00000&  \\
41292.965000 &  -78.000000 $\pm$ 8.500000 & 94.600000   $\pm$  13.00000&  \\
41343.754000 &  131.700000 $\pm$ 8.500000 & -85.100000  $\pm$  13.00000&  \\
41345.683000 &  -62.900000 $\pm$ 8.500000 & 73.600000   $\pm$  13.00000&  \\
41348.689000 &  110.800000 $\pm$ 8.500000 & -115.000000 $\pm$  13.00000&  \\
             &                        &                       & This paper\\
50386.658000 &  -68.060000 $\pm$ 2.000000 & 71.317000   $\pm$  2.010000&  \\
50388.689000 &  65.666000  $\pm$ 2.000000 & -61.937000  $\pm$  2.000000&  \\
50389.564000 &  106.155000 $\pm$ 2.000000 & -101.269000 $\pm$  2.010000&  \\
50389.682000 &  114.474000 $\pm$ 2.000000 & -106.669000 $\pm$  2.010000&  \\
50389.691000 &  111.486000 $\pm$ 2.000000 & -104.108000 $\pm$  2.010000&  \\
50419.603000 &  45.879000  $\pm$ 2.000000 & -39.742000  $\pm$  2.010000&  \\
50420.679000 &  88.566000  $\pm$ 4.000000 & -82.566000  $\pm$  4.000000&  \\
50803.603000 &  -93.102000 $\pm$ 2.000000 & 94.896000   $\pm$  2.010000&  \\
50803.662000 &  -114.114000$\pm$ 2.000000 & 117.833000  $\pm$  2.010000&  \\
50804.466000 &  -111.310000$\pm$ 2.000000 & 115.248000  $\pm$  2.010000&  \\
50804.512000 &  -122.679000$\pm$ 2.000000 & 122.554000  $\pm$  2.010000&  \\
50804.554000 &  -111.279000$\pm$ 2.000000 & 114.730000  $\pm$  2.010000&  \\
50804.613000 &  -87.933000 $\pm$ 2.000000 & 88.354000   $\pm$  2.010000&  \\
50804.655000 &  -48.587000 $\pm$ 2.000000 & 55.073000   $\pm$  2.010000&  \\
50804.702000 &  -12.103000 $\pm$ 5.000000 & 19.282000   $\pm$  5.000000&  \\
50837.280000 &  -8.535000  $\pm$ 5.000000 & 15.242000   $\pm$  5.000000&  \\
50837.344000 &  53.059000  $\pm$ 2.000000 & -48.892000  $\pm$  2.010000&  \\
50837.441000 &  114.356000 $\pm$ 2.000000 & -110.317000 $\pm$  2.010000&  \\
50837.482000 &  123.220000 $\pm$ 2.000000 & -120.601000 $\pm$  2.010000&  \\
50837.544000 &  111.180000 $\pm$ 2.000000 & -107.552000 $\pm$  2.010000&  \\
50837.573000 &  99.324000  $\pm$ 2.000000 & -93.619000  $\pm$  2.010000&  \\
50837.607000 &  76.596000  $\pm$ 2.000000 & -72.885000  $\pm$  2.010000&  \\
50837.671000 &  27.926000  $\pm$ 4.000000 & -21.195000  $\pm$  4.000000&  \\
50838.279000 &  121.265000 $\pm$ 2.000000 & -119.862000 $\pm$  2.010000&  \\
50838.323000 &  120.991000 $\pm$ 2.000000 & -119.629000 $\pm$  2.010000&  \\
50838.370000 &  106.760000 $\pm$ 2.000000 & -101.976000 $\pm$  2.010000&  \\
50838.402000 &  91.200000  $\pm$ 2.000000 & -86.156000  $\pm$  2.010000&  \\
50838.441000 &  56.558000  $\pm$ 2.000000 & -53.281000  $\pm$  2.010000&  \\
50838.475000 &  34.583000  $\pm$ 3.000000 & -29.543000  $\pm$  3.000000&  \\
50838.571000 &  -55.664000 $\pm$ 2.000000 & 57.430000   $\pm$  2.010000&  \\
50838.601000 &  -77.094000 $\pm$ 2.000000 & 85.060000   $\pm$  2.010000&  \\
50839.422000 &  -85.943000 $\pm$ 2.000000 & 90.506000   $\pm$  2.010000&  \\
50839.454000 &  -101.098000$\pm$ 2.000000 & 105.197000  $\pm$  2.010000&  \\
50839.501000 &  -119.696000$\pm$ 2.000000 & 121.266000  $\pm$  2.010000&  \\
50839.557000 &  -117.400000$\pm$ 2.000000 & 118.909000  $\pm$  2.010000&  \\
50839.618000 &  -91.110000 $\pm$ 2.000000 & 93.943000   $\pm$  2.010000&  \\
50839.681000 &  -39.312000 $\pm$ 2.000000 & 45.560000   $\pm$  2.010000&  \\
\hline
\end{tabular}  
\caption{Radial-velocity measurements of YYGem. 
To be published in electronic form only.}
\label{ind_meas2}
\end{table*}

\begin{table*}
\begin{tabular}{|l|l|l|l|} 
\multicolumn{4}{c}{GJ2069A}\\\hline        
Julian Day   &  V$_A$               &     V$_B$               & References \\
2400000+     & (in km/s)            & (in km/s)               &  \\
\hline\hline                                                                
             &                      &                         &  this paper\\
50102.528000 &  72.363000 $\pm$ 0.387000 & -70.039000 $\pm$  0.671000 & \\ 
50180.417000 &  55.762000 $\pm$ 0.387000 & -51.727000 $\pm$  0.671000 & \\ 
50181.389000 &  -62.507000 $\pm$ 0.387000 & 76.540000 $\pm$  0.671000 & \\ 
50181.442000 &  -63.163000 $\pm$ 0.387000 & 78.303000 $\pm$  0.671000 & \\ 
50182.323000 &  26.906000 $\pm$ 0.387000 & -20.842000 $\pm$  0.671000 & \\ 
50182.369000 &  33.245000 $\pm$ 0.387000 & -26.164000 $\pm$  0.671000 & \\ 
50182.430000 &  41.317000 $\pm$ 0.387000 & -37.358000 $\pm$  0.671000 & \\ 
50182.476000 &  46.551000 $\pm$ 0.387000 & -42.814000 $\pm$  0.671000 & \\ 
50183.384000 &  31.521000 $\pm$ 0.387000 & -23.989000 $\pm$  0.671000 & \\ 
50183.427000 &  25.212000 $\pm$ 0.387000 & -18.244000 $\pm$  0.671000 & \\ 
50183.491000 &  16.000000 $\pm$ 0.387000 & -8.064000 $\pm$  0.671000  & \\ 
50228.354000 &  -53.128000 $\pm$ 0.387000 & 67.222000 $\pm$  0.671000 & \\ 
50231.357000 &  -62.646000 $\pm$ 0.387000 & 79.392000 $\pm$  0.671000 & \\ 
50389.667000 &  -46.490000 $\pm$ 0.200000 & 59.689000 $\pm$  0.300000 & \\ 
50525.433000 &  -50.097000 $\pm$ 0.200000 & 63.745000 $\pm$  0.300000 & \\ 
50921.344000 &  -61.063000 $\pm$ 0.200000 & 75.378000 $\pm$  0.300000 & \\ 
50922.330000 &  29.951000 $\pm$ 0.200000 & -23.427000 $\pm$  0.300000 & \\ 
50923.332000 &  36.079000 $\pm$ 0.200000 & -30.010000 $\pm$  0.300000 & \\ 
\hline
\end{tabular}  
\caption{Radial-velocity of measurements of GJ~2069A. 
To be published in electronic form only.}
\label{ind_meas3}
\end{table*}

\begin{table*}
\begin{tabular}{|l|l|l|l|l|} 
\multicolumn{5}{c}{Gl644ABC}\\\hline     
Julian Day   &  V$_A$               &     V$_B$               &    V$_C$     & References \\
2400000+     & (in km/s)            & (in km/s)               & (in km/s)    &            \\ \hline\hline      
             &                      &                             &           &this paper\\
50182.632800 & 10.129 $\pm$ 0.2     & 33.141 $\pm$ 0.1       &  0.882 $\pm$ 0.1 &\\
50228.466500 & 11.377 $\pm$ 0.3     &  3.207 $\pm$ 0.3       & 32.846 $\pm$ 0.1 &\\
50522.682700 & 18.770 $\pm$ 0.1     & -0.128 $\pm$ 0.1       & 26.453 $\pm$ 0.5 &\\
50523.677200 & 18.775 $\pm$ 0.1     & 28.264 $\pm$ 0.1       & -4.916 $\pm$ 0.1 &\\
50525.676200 & 18.770 $\pm$ 0.1     &  0.709 $\pm$ 0.1       & 25.831 $\pm$ 1.0 &\\
50577.575700 & 16.250 $\pm$ 0.2     & 25.700 $\pm$ 0.2       &  0.776 $\pm$ 0.1 &\\
50578.575200 & 16.253 $\pm$ 0.1     & -2.642 $\pm$ 0.1       & 32.580 $\pm$ 0.1 &\\
50641.367100 & 13.100 $\pm$ 0.1     &  5.449 $\pm$ 0.1       & 28.058 $\pm$ 0.1 &\\
50642.363200 & 13.138 $\pm$ 0.1     & 32.297 $\pm$ 0.1       & -1.761 $\pm$ 0.1 &\\    
50642.450100 & 13.170 $\pm$ 0.1     & 32.563 $\pm$ 0.1       & -2.097 $\pm$ 0.1 &\\
50643.370600 & 13.100 $\pm$ 0.2     & 10.585 $\pm$ 0.2       & 21.862 $\pm$ 0.3 &\\
50644.383200 & 13.100 $\pm$ 0.1     &  6.955 $\pm$ 0.1       & 26.544 $\pm$ 0.12 &\\
50645.402500 & 13.007 $\pm$ 0.1     & 32.658 $\pm$ 0.1       & -2.013 $\pm$ 0.1 &\\
50647.489500 & 13.100 $\pm$ 0.1     & 11.551 $\pm$ 0.3       & 21.004 $\pm$ 0.3 &\\
50673.342200 & 11.840 $\pm$ 0.2     &  2.608 $\pm$ 0.2       & 33.383 $\pm$ 0.1 &\\
50674.375400 & 11.840 $\pm$ 0.2     & 18.198 $\pm$ 0.7       & 15.992 $\pm$ 0.7 &\\
50922.626000 & 14.963 $\pm$ 0.1     & -2.053 $\pm$ 0.1       & 33.717 $\pm$ 0.15 &\\
50983.445800 & 18.133 $\pm$ 0.1     & 29.089 $\pm$ 0.15      & -5.269 $\pm$ 0.1 &\\
50986.460400 & 18.221 $\pm$ 0.1     & 29.203 $\pm$ 0.1       & -5.425 $\pm$ 0.1 &\\
51017.372300 & 19.600 $\pm$ 0.25    & -3.340 $\pm$ 0.1       & 28.400 $\pm$ 1.0 &\\
51026.368200 & 19.428 $\pm$ 0.15    & -4.932 $\pm$ 0.1       & 28.943 $\pm$ 0.6 &\\  \hline
\end{tabular}  
\caption{Radial-velocity measurements of Gl~644ABC. 
To be published in electronic form only.}
\label{ind_meas4}
\end{table*}

\begin{table*}
\begin{tabular}{|l|l|l|l|} 
\multicolumn{4}{c}{Gl747AB}\\\hline        
Julian Day   &  V$_A$               &     V$_B$               & References \\
2400000+     & (in km/s)            & (in km/s)               & \\
\hline\hline                                                                  
             &                      &                         & this paper\\
49977.364600 &  -50.669 $\pm$ 0.300 & -44.148 $\pm$ 0.300 & \\ 
50316.358000 &  -43.284 $\pm$ 0.028 & -51.687 $\pm$ 0.033 & \\ 
50389.276800 &  -41.424 $\pm$ 0.040 & -53.637 $\pm$ 0.048 & \\ 
50419.282000 &  -40.684 $\pm$ 0.072 & -54.347 $\pm$ 0.089 & \\ 
50525.626700 &  -39.752 $\pm$ 0.037 & -55.393 $\pm$ 0.044 & \\ 
50576.609400 &  -39.877 $\pm$ 0.056 & -55.123 $\pm$ 0.068 & \\ 
50641.514600 &  -40.692 $\pm$ 0.025 & -54.390 $\pm$ 0.032 & \\ 
50673.400000 &  -41.078 $\pm$ 0.022 & -53.917 $\pm$ 0.025 & \\ 
50747.277800 &  -42.416 $\pm$ 0.040 & -52.784 $\pm$ 0.040 & \\ 
50923.634600 &  -46.246 $\pm$ 0.500 & -49.668 $\pm$ 0.500 & \\
51244.681500 &  -49.145 $\pm$ 0.500 & -45.191 $\pm$ 0.500 & \\
51452.415900 &  -51.065 $\pm$ 0.040 & -43.414 $\pm$ 0.050 & \\
51502.257600 &  -51.429 $\pm$ 0.050 & -43.378 $\pm$ 0.050 & \\
51648.640100 &  -51.951 $\pm$ 0.050 & -42.621 $\pm$ 0.050 & \\
51717.538800 &  -51.956 $\pm$ 0.050 & -42.274 $\pm$ 0.050 & \\ \hline
\end{tabular}   
\caption{Radial-velocity measurements of Gl~747. 
To be published in electronic form only.}
\label{ind_meas5}
\end{table*}

\begin{table*}
\begin{tabular}{|l|l|l|l|} 
\multicolumn{4}{c}{Gl831AB}\\\hline        
Julian Day   &  V$_A$               &     V$_B$               & References \\
2400000+     & (in km/s)            & (in km/s)               & \\
\hline\hline      
             &                      &                         & this paper \\
49977.4017   &  -58.778 $\pm$ .100  &  -53.553 $\pm$ 1.000    & \\
50313.4914   &  -55.660 $\pm$ .500  &                         & \\
50386.2936   &  -51.831 $\pm$ .100  &  -61.573 $\pm$ 1.000    & \\
50420.2883   &  -49.608 $\pm$ .100  &  -69.171 $\pm$ 2.000    & \\
50642.5675   &  -57.856 $\pm$ .100  &  -53.964 $\pm$ 1.000    & \\
50673.5155   &  -58.450 $\pm$ .100  &  -52.600 $\pm$ 1.000    & \\
50746.2883   &  -59.194 $\pm$ .100  &  -52.852 $\pm$ 1.000    & \\
50983.5875   &  -56.364 $\pm$ .500  &                         & \\
51019.5585   &  -55.788 $\pm$ .500  &                         & \\
51109.3370   &  -50.642 $\pm$ .050  &  -65.978 $\pm$  .500    & \\
51162.2405   &  -48.499 $\pm$ .100  &  -72.198 $\pm$ 1.000    & \\
51163.2380   &  -48.772 $\pm$ .100  &  -70.663 $\pm$ 1.000    & \\
51364.5670   &  -57.635 $\pm$ .500  &                         & \\
51396.5163   &  -58.358 $\pm$ .500  &                         & \\
51451.3961   &  -58.683 $\pm$ .500  &                         & \\ \hline
\hline
\end{tabular}  
\caption{Radial-velocity measurements of Gl~831. 
To be published in electronic form only.}
\label{ind_meas6}
\end{table*}

\begin{table*}
\begin{tabular}{|l|l|l|l|l|} 
\multicolumn{5}{c}{Gl866ABC}\\\hline      
Julian Day   &  V$_A$               &     V$_B$               &    V$_C$    & References \\
2400000+     & (in km/s)            & (in km/s)               & (in km/s) & \\ \hline\hline      
             &                      &                         &            & Delfosse et al. 1999b \\
50019.325      &   -25.004  $\pm$  .100  &     -53.301 $\pm$   .080  &       -77.626 $\pm$   .300 & \\  
50313.543      &   -79.960  $\pm$  .100  &     -46.680 $\pm$   .120  & & \\
50386.336      &   -38.987  $\pm$  .100  & & & \\
50389.319      &   -76.516  $\pm$  .100  &     -50.815 $\pm$   .080  &       -15.680 $\pm$   .500 & \\
50418.269      &   -46.732  $\pm$  .700  & & & \\
50419.244      &   -81.252  $\pm$  .100  &     -52.148 $\pm$   .100  & & \\
50419.324      &   -80.921  $\pm$  .100  &     -51.979 $\pm$   .080  & & \\
50420.249      &   -47.177  $\pm$ 1.000  & & & \\
50641.595      &   -40.094  $\pm$  .100  &     -56.016 $\pm$   .070  & & \\
50642.594      &   -78.992  $\pm$  .300  &     -55.744 $\pm$   .120  & & \\
50643.596      &   -47.563  $\pm$  .300  & & & \\
50644.594      &   -14.713  $\pm$  .100  &     -55.855 $\pm$   .070  & & \\
50645.598      &   -50.656  $\pm$ 1.000  & & & \\
50646.592      &   -78.420  $\pm$  .100  &     -55.849 $\pm$   .100  &         -7.567 $\pm$   .400 & \\
50647.605      &   -35.530  $\pm$  .300  & & & \\
50672.583      &   -72.725  $\pm$  .110  &     -56.314 $\pm$   .110  & & \\
50673.471      &   -67.182  $\pm$  .200  &                           &        -19.782 $\pm$   .300 & \\
50674.617      &   -17.102  $\pm$  .100  &     -56.341 $\pm$   .060  & & \\
50976.890      &   -54.268  $\pm$ 1.000  &     -37.044 $\pm$   .100  & & \\
50998.903      &   -85.950  $\pm$  .300  &     -35.550 $\pm$   .100  &        -23.400 $\pm$   .500 & \\
51020.847      &   -80.600  $\pm$  .500  &     -35.700 $\pm$   .200  &        -30.000 $\pm$   .500 & \\
51068.673      &   -23.183  $\pm$  .100  &     -40.454 $\pm$   .100  & & \\
51017.580      &   -89.724  $\pm$  .080  &     -35.289 $\pm$   .100  & & \\
51108.353      &   -85.172  $\pm$  .050  &     -44.530 $\pm$   .050  &        -13.116 $\pm$   .200 & \\
51159.238      &   -22.522  $\pm$  .200  &     -48.651 $\pm$   .200  & & \\
51161.231      &   -81.280  $\pm$  .200  &     -48.331 $\pm$   .200  & & \\
51363.594      &   -22.324  $\pm$  .070  &     -55.070 $\pm$   .090  &        -79.100 $\pm$   .310 & \\ \hline
\end{tabular}  
\caption{Radial-velocity measurements of Gl~866ABC. 
To be published in electronic form only.}
\label{ind_meas7}
\end{table*}

\begin{table*}
\begin{tabular}{|l|l|l|ll|} 
\multicolumn{5}{c}{Gl234AB}\\\hline                                                             
Julian Day   & $\rho$     & $\theta$  & Method & References \\ 
 2400000+    &(arc sec)   & (degree)  &    & \\\hline \hline   
35189.500 &1.190 $\pm$  .050&   36.0 $\pm$    5.0  & photograph & Probst et al. 1977\\ 
35545.500 &1.160 $\pm$  .050&   43.2 $\pm$    5.0  & visual & Probst et al. 1977\\ 
35942.500 &1.220 $\pm$  .050&   51.9 $\pm$    5.0  & visual & Probst et al. 1977\\
42063.500 &1.270 $\pm$   .050&   53.0 $\pm$   40.0  & photoelec & Probst et al. 1977 \\
46778.500 &1.110 $\pm$   .110&   22.6 $\pm$    2.0  & speckle & Coppenbarger et al. 1994\\
48583.500 &1.190 $\pm$   .048&   66.0 $\pm$    2.0  & speckle & Coppenbarger et al. 1994\\
48673.500 &1.100 $\pm$   .044&   66.3 $\pm$    2.0  & speckle & Coppenbarger et al. 1994\\
51506.080 & .457 $\pm$   .001&  270.4 $\pm$     .1  & A.O. & this paper\\      
51592.900 & .439 $\pm$   .0005& 287.4 $\pm$     .1  & A.O. & this paper\\ 
51655.315 & .444 $\pm$   .001&  299.7 $\pm$     .1  & A.O. & this paper\\ \hline
          &  Position Angle  &                     Projected separations & & \\ \hline      
46368.500 &  1.040 $\pm$    .100  &  .0  & speckle & Coppenbarger et al. 1994\\       
47107.500 &  1.120 $\pm$    .110  &  .0  & speckle & Coppenbarger et al. 1994\\       
47198.500 &   .980 $\pm$    .090  &90.0  & speckle & Coppenbarger et al. 1994\\       
47576.500 &   .810 $\pm$    .080  &90.0  & speckle & Coppenbarger et al. 1994\\       
\hline
\end{tabular}  
\caption{Angular separation measurements for Gl~234AB. 
To be published in electronic form only.}
\label{ind_meas8}
\end{table*}

\begin{table*}
\begin{tabular}{|l|l|l|ll|} 
\multicolumn{5}{c}{Gl644A-BC}    \\\hline     
Julian Day  & $\rho$     & $\theta$  & Type & References   \\                                       
2400000+    &(arc sec)   & (degree)  &      &    \\\hline \hline         
45863.690&  .230$\pm$  .020 &  115.4 $\pm$  5.0  & speckle & Al Shukri et al. 1996 \\
45864.603&  .243$\pm$  .030 &  115.1 $\pm$  5.0  & speckle & Al Shukri et al. 1996 \\
46592.937&  .197$\pm$  .011 &   48.5 $\pm$  2.5  & speckle & Blazit et al. 1987 \\
46593.960&  .208$\pm$  .006 &   41.3 $\pm$  1.5  & speckle & Blazit et al. 1987 \\
46934.452&  .242$\pm$  .010 &  206.9 $\pm$  2.0  & speckle & Balega et al. 1989 \\
47260.554&  .213$\pm$  .010 &   16.9 $\pm$  2.0  & speckle & Balega et al. 1991 \\
47676.373&  .240$\pm$  .015 &  143.8 $\pm$  5.0  & speckle & Balega et al. 1994 \\
48378.066&  .211$\pm$  .010 &  102.7 $\pm$  3.0  & speckle & Hartkopf et al. 1994 \\
51017.500&  .216$\pm$  .002 &   21.9 $\pm$   .3  & A.O. & this paper \\
51414.780&  .231.2$\pm$  .0005 &  161.5 $\pm$   .1  & A.O. & this paper \\    
51594.128&  .207.7$\pm$  .0005 &   53.6 $\pm$   .1  & A.O. & this paper \\
51650.080&  .213$\pm$  .001 &   18.3 $\pm$   .2  & A.O. & this paper \\
51717.956&  .216$\pm$  .004 &  340.3 $\pm$  1.0  & A.O. & this paper \\ \hline
\end{tabular}  
\caption{Angular separation measurements for Gl644A-BC. 
To be published in electronic form only.}
\label{ind_meas9}
\end{table*}

\begin{table*}
\begin{tabular}{|l|l|l|ll|} 
\multicolumn{5}{c}{Gl747AB}    \\\hline        
Julian Day   &  $\rho$     & $\theta$  & Type & References   \\                                      
2400000+     &(arc sec)   & (degree)  &      &                 \\ \hline \hline        
46596.005  & .148 $\pm$ .010  & 110.6 $\pm$  3.9    & speckle & Blazit et al. 1987 \\
46580.446  & .147 $\pm$ .010  & 110.6 $\pm$  3.9    & speckle & Balega et al. 1989 \\
46669.309  & .089 $\pm$ .010  & 117.4 $\pm$  6.4    & speckle & Balega et al. 1989 \\
46934.525  & .141 $\pm$ .010  & 233.0  $\pm$ 4.1    & speckle & Balega et al. 1989 \\
51413.815  & .304  $\pm$.004  & 256.1 $\pm$   .1    & A.O. & this paper \\
51414.815  & .298  $\pm$.001  & 255.9 $\pm$   .1    & A.O. & this paper \\
51508.191  & .331 $\pm$ .001  & 258.6 $\pm$   .1    & A.O. & this paper \\
51645.119  & .3542 $\pm$ .0004& 262.3 $\pm$   .1    & A.O. & this paper \\
\hline      
 \end{tabular}  
\caption{Angular separation measuremens for Gl~747AB. 
To be published in electronic form only.}
\label{ind_meas10}
\end{table*}

\begin{table*}
\begin{tabular}{|l|l|l|ll|} 
\multicolumn{5}{c}{Gl831AB}    \\\hline        
Julian Day   &  $\rho$     & $\theta$  & Type & References  \\                                      
2400000+     & (arc sec)   & (degree)  &      &                \\\hline \hline        
51414.925  &.187 $\pm$ .002  & 134.4   $\pm$ .4   & A.O. & this paper \\
51507.700  &.198 $\pm$ .001  & 149.4   $\pm$ .2   & A.O. & this paper \\
51715.091  &.109 $\pm$ .004  & 203.1   $\pm$ .4   & A.O. & this paper \\
\hline      
 \end{tabular}  
\caption{Angular separation measurements for Gl~831AB. 
To be published in electronic form only.}
\label{ind_meas11}
\end{table*}

\begin{table*}
\begin{tabular}{|l|l|l|ll|} 
\multicolumn{5}{c}{Gl866AC-B}   \\ \hline  
   Julian Day      &      X        &        Y       & Type & References\\      
2400000+     &        (arc sec)             &       (arc sec)          & & \\ \hline \hline
46381.500&   .400 $\pm$ .020 &    .000 $\pm$ .020 &     speckle             & Leinert et al. 1990  \\
46385.500&   .390 $\pm$ .020 &    .000 $\pm$ .020 &     speckle             & Leinert et al. 1990  \\
46594.500&   .435 $\pm$ .010 &   -.165 $\pm$ .010 &     speckle             & Leinert et al. 1990  \\
46603.500&   .440 $\pm$ .010 &   -.165 $\pm$ .010 &     speckle             & Leinert et al. 1990  \\
46686.500&   .275 $\pm$ .020 &   -.200 $\pm$ .020 &     speckle             & Leinert et al. 1990  \\
46690.500&   .280 $\pm$ .020 &   -.175 $\pm$ .020 &     speckle             & Leinert et al. 1990  \\
46952.500&  -.105 $\pm$ .025 &    .105 $\pm$ .025 &     speckle             & Leinert et al. 1990  \\
46954.500&                   &    .110 $\pm$ .020 &     speckle             & Leinert et al. 1990  \\
46962.500&  -.085 $\pm$ .010 &    .130 $\pm$ .010 &     speckle             & Leinert et al. 1990  \\
47044.500&   .120 $\pm$ .020 &    .085 $\pm$ .020 &     speckle             & Leinert et al. 1990  \\
47048.500&   .125 $\pm$ .020 &    .100 $\pm$ .020 &     speckle             & Leinert et al. 1990  \\
47107.500&   .290 $\pm$ .020 &                    &     speckle             & Leinert et al. 1990  \\ 
47165.500&   .375 $\pm$ .020 &    .020 $\pm$ .020 &     speckle             & Leinert et al. 1990  \\
47339.500&   .445 $\pm$ .030 &   -.085 $\pm$ .030 &     speckle             & Leinert et al. 1990  \\
47410.500&   .440 $\pm$ .025 &   -.175 $\pm$ .025 &     speckle             & Leinert et al. 1990  \\ \hline
        &  $\rho$     & $\theta$  & & \\
     &(arc sec)   & (degree)  &      &               \\ \hline
48136.518&  .481 $\pm$  .018  & 348.0$\pm$   2.2   &  speckle               &  Woitas et al. 2000 \\
48230.495&  .463 $\pm$  .019  & 339.8$\pm$   2.4   &  speckle               &  Woitas et al. 2000 \\
48559.482&  .200 $\pm$  .013  & 155.5$\pm$   3.8   &  speckle               &  Woitas et al. 2000 \\
48759.502&  .244 $\pm$  .006  &  15.3$\pm$    .5   &  speckle               &  Woitas et al. 2000 \\
48910.493&  .449 $\pm$  .011  & 350.8$\pm$   1.5   &  speckle               &  Woitas et al. 2000 \\
48998.471&  .480 $\pm$  .012  & 348.6$\pm$   1.4   &  speckle               &  Woitas et al. 2000 \\
49198.514&  .259 $\pm$  .011  & 319.3$\pm$   1.2   &  speckle               &  Woitas et al. 2000 \\
49263.491&  .105 $\pm$  .004  & 277.7$\pm$    .5   &  speckle               &  Woitas et al. 2000 \\
49474.505&  .126 $\pm$  .004  &  74.9$\pm$   2.7   &  speckle               &  Woitas et al. 2000 \\
49610.484&  .327 $\pm$  .004  &   5.9$\pm$    .4   &  speckle               &  Woitas et al. 2000 \\
49699.494&  .434 $\pm$  .004  & 354.0$\pm$    .2   &  speckle               &  Woitas et al. 2000 \\
49700.516&  .439 $\pm$  .004  & 354.3$\pm$    .3   &  speckle               &  Woitas et al. 2000 \\
49908.462&  .439 $\pm$  .004  & 337.3$\pm$    .3   &  speckle               &  Woitas et al. 2000 \\
49946.521&  .392 $\pm$  .004  & 333.1$\pm$    .3   &  speckle               &  Woitas et al. 2000 \\
50243.513&  .156 $\pm$  .003  & 131.6$\pm$   1.1   &  HST FGS3              &  Woitas et al. 2000 \\
50354.473&  .188 $\pm$  .004  &  28.0$\pm$    .5   &  speckle               &  Woitas et al. 2000 \\
50625.500&  .488 $\pm$  .007  & 345.8$\pm$   1.7   &  HST  Planetary Camera & Schroeder et al. 2000 \\
50656.500&  .486 $\pm$  .007  & 343.8$\pm$   1.7   &  HST  Planetary Camera & Schroeder et al. 2000 \\
50686.506&  .485 $\pm$  .004  & 340.2$\pm$    .2   &  speckle               &  Woitas et al. 2000 \\
50769.489&  .391 $\pm$  .004  & 333.3$\pm$    .1   &  speckle               &  Woitas et al. 2000 \\
50941.495&  .104 $\pm$  .007  & 214.8$\pm$    .3   &  speckle               &  Woitas et al. 2000 \\
51017.500&  .200 $\pm$ .003  & 155.9 $\pm$   .9    &  A.O.                  & this paper \\
51017.500&  .198 $\pm$  .003  & 156.2$\pm$    .9   &  A.O.                  & this paper \\
51097.490&  .133 $\pm$  .004  & 101.2$\pm$    .2   &  speckle               & Woitas et al. 2000 \\     
51414.925&  .480 $\pm$  .001  & 348.4$\pm$    .1   &  A.O.                  & this paper \\
51507.725&  .473 $\pm$  .001  & 341.0$\pm$    .1   &  A.O.                  & this paper \\ \hline
\end{tabular}                                                                             
\caption{angular measurements for Gl866AC-B. 
To be published in electronic form only.}
\label{ind_meas12}
\end{table*}

\end{document}